\documentclass[a4paper,aps,prd,superscriptaddress,nofootinbib,preprintnumbers]{revtex4}
\usepackage{amsmath,multirow}
\usepackage{graphicx,tabularx,slashed}
\usepackage{url}
\usepackage{hyperref}

\usepackage{color}

\def\tablename{Table}
\def\figurename{Figure}

 \usepackage{bbold}
\newcommand{\sw}{\ensuremath{s_W}}

\newcommand{\swd}{\ensuremath{s_W^2}}

\newcommand{\sql}{\ensuremath{\tilde{q}_L}}
\newcommand{\sqr}{\ensuremath{\tilde{q}_R}}
\newcommand{\sq}{\ensuremath{\tilde{q}}}

\newcommand{\mssqo}{\ensuremath{m^2_{\sq_1}}}
\newcommand{\mssqt}{\ensuremath{m^2_{\sq_2}}}

\newcommand{\rtoo}{\ensuremath{R^{\tilde{t}}_{11}}}
\newcommand{\rtot}{\ensuremath{R^{\tilde{t}}_{12}}}
\newcommand{\rtto}{\ensuremath{R^{\tilde{t}}_{21}}}
\newcommand{\rttt}{\ensuremath{R^{\tilde{t}}_{22}}}

\newcommand{\retildehat}{\ensuremath{\Re\,\text{e} \hat{\Sigma}}}
\newcommand{\retilde}{\ensuremath{\Re\,\text{e} \Sigma}}

\newcommand{\gtchalo}{\ensuremath{g_{\stop\chai\,b, L}^{(1)}}}
\newcommand{\gtchalt}{\ensuremath{g_{\stop\chai\,b, L}^{(2)}}}
\newcommand{\gtcharo}{\ensuremath{g_{\stop\chai\,b, R}^{(1)}}}
\newcommand{\gtchart}{\ensuremath{g_{\stop\chai\,b, R}^{(2)}}}
\newcommand{\gtchala}{\ensuremath{g_{\stop\chai\,b, L}^{(a)}}}
\newcommand{\gtchara}{\ensuremath{g_{\stop\chai\,b, R}^{(a)}}}
\newcommand{\gptchalo}{\ensuremath{g_{\stop^*\chaip\,b, L}^{(1)}}}
\newcommand{\gptchalt}{\ensuremath{g_{\stop^*\chaip\,b, L}^{(2)}}}
\newcommand{\gptcharo}{\ensuremath{g_{\stop^*\chaip\,b, R}^{(1)}}}
\newcommand{\gptchart}{\ensuremath{g_{\stop^*\chaip\,b, R}^{(2)}}}
\newcommand{\gptchala}{\ensuremath{g_{\stop^*\chaip\,b, L}^{(a)}}}
\newcommand{\gptchara}{\ensuremath{g_{\stop^*\chaip\,b, R}^{(a)}}}
\newcommand{\myrbox}[1]{\parbox{4.0cm}{#1}}

\newcommand{\sqt}{\ensuremath{\tilde{t}}}
\newcommand{\sqb}{\ensuremath{\tilde{b}}}
\newcommand{\rotsq}{\ensuremath{R^{\sq}}}

\newcommand{\lag}{\mathcal{L}}

\newcommand{\qqqquad}{\qquad \qquad \qquad}

\newcommand{\msbar}{\ensuremath{\overline{\text{MS}}}}

\newcommand{\go}{\tilde{g}}
\newcommand{\mg}{{\sc MadGolem\,}}

\renewcommand{\stop}{\ensuremath{\tilde{t}}}
\newcommand{\sbot}{\ensuremath{\tilde{b}}}
\newcommand{\stone}{\ensuremath{\tilde{t}_1}}

\newcommand{\neutone}{\ensuremath{\tilde{\chi}^0_1}}

\newcommand{\chaone}{\ensuremath{\tilde{\chi}^{\pm}_1}}
\newcommand{\chai}{\ensuremath{\tilde{\chi}^{-}_i}}
\newcommand{\chaip}{\ensuremath{\tilde{\chi}^{+}_i}}

\def\slashchar#1{\setbox0=\hbox{$#1$}           % set a box for #1
   \dimen0=\wd0                                 % and get its size
   \setbox1=\hbox{/} \dimen1=\wd1               % get size of /
   \ifdim\dimen0>\dimen1                        % #1 is bigger
      \rlap{\hbox to \dimen0{\hfil/\hfil}}      % so center / in box
      #1                                        % and print #1
   \else                                        % / is bigger
      \rlap{\hbox to \dimen1{\hfil$#1$\hfil}}   % so center #1
      /                                         % and print /
   \fi}

\def\eg{{\sl e.g.} \,}

\def\etal{{\sl et al} \,}
\begin{document}

\date{\today}

\title{Automated third generation squark production to next-to-leading order}

\preprint{IPPP/14/66}
\preprint{DCPT/14/132}

\author{Dorival Gon\c{c}alves}
\affiliation{Institute for Particle Physics Phenomenology, Department
of Physics, Durham University, United Kingdom}

\author{David L\'opez-Val}
\affiliation{Centre for Cosmology, Particle Physics \& Phenomenology CP3, Universit\'e catholique de Louvain,
Belgium}

\author{Kentarou Mawatari}
\affiliation{Theoretische Natuurkunde and IIHE/ELEM, Vrije Universiteit Brussel, Belgium \\
             and International Solvay Institutes, Brussels, Belgium}

\author{Tilman Plehn}
\affiliation{Institut f\"ur Theoretische Physik, Universit\"at Heidelberg, Germany}

\begin{abstract}
If light--flavor squarks and gluinos are indeed heavy and chargino
pair production is plagued with overwhelming backgrounds, pair
production and associated production of stops and charginos will
become the key signatures in the upcoming LHC runs.  We present fully
automated next-to-leading order predictions for heavy flavor squark
production at the LHC, including stop--chargino associated production,
based on \textsc{MadGolem}.  We compute the total and differential NLO
rates for a variety of MSSM scenarios with a light third
generation. We focus on theoretical uncertainties on differential
rates, including a comparison to multi-jet merging and the impact of
bottom parton densities for stop--chargino associated production. We
find that for the associated production channel the rates can reach
tens of femtobarns, but the scale dependence does not provide a
conservative estimate of the theoretical uncertainties.
\end{abstract}

\maketitle
\tableofcontents

\newpage
%%%%%%%%%%%%%%%%%%%%%%%%%%%%%%%%%%%%%%%%%%%%%%%%%%%%%%%%%%%%%%%%%%%%%%%%
\section{Introduction}

After the discovery of a light, likely fundamental Higgs scalar,
supersymmetry (SUSY) as a solution to the hierarchy problem is still
among the most cherished paradigms for physics beyond the Standard
Model (SM)~\cite{Morrissey:2009tf}.  In relation to experimental
results supersymmetry provides dark matter candidates and a pathway
towards gauge coupling unification.  Notwithstanding, TeV-scale SUSY
is in growing tension with the Run-I LHC results~\cite{susy_fits}.
Inclusive searches are largely governed by light--flavor squarks and
gluino production and provide exclusion limits for example of the kind
$m_{\sq}\simeq m_{\go}\simeq 1.5$~TeV or $m_{\go}\simeq 1.2$~TeV for
$m_{\go}\ll m_{\sq}$ in the squark--gluino mass
plane~\cite{searches-gen,atlassite,cmssite}. In comparison, the limits
on direct stop pair production as well as direct neutralino and
chargino production hardly affect the supersymmetric parameter
space. Be as it may, the inclusive search strategies do not cover the
whole MSSM parameter space and motivate more flexible
approaches for the upcoming LHC runs.\bigskip

In particular third--generation squarks with their direct link to the
hierarchy problem are being searched for in dedicated analyses.
Renormalization group effects as well as squark mixing tend to make
one top squark especially
light~\cite{nsusy,nsusy-benchmarks,Buchmueller:2013exa,Brummer:2012ns,Han:2013mga}.
To avoid the gauge link between left--handed stops and sbottoms such a
light stop will typically be right--handed.  In dark matter models a
light top squark can co-annihilate with a bino--like neutralino,
giving the correct relic density for a $\mathcal{O}(10)$~GeV mass
splitting~\cite{stop_coann}.  In the context of electroweak
baryogenesis, light stops may allow for the required first-order phase
transition and pull down the critical temperature to protect the
baryon asymmetry from washout~\cite{Morrissey:2012db}. All of this points 
to scenarios, where at least one third generation squark 
and the electroweak gauginos and Higgsinos are relatively light, while 
light--flavor squarks and gluinos are too heavy to be observed at the LHC
during an early 14~TeV run.\bigskip

The LHC results from the 7~TeV and 8~TeV runs translate into exclusion
limits in the $\tilde{t}_1-\neutone$ mass plane. Assuming R-parity
conservation with the lightest stop as the next-to-lightest
supersymmetric partner (NLSP) the decay mode $\stone\to t\neutone$
requires $m_{\stone}\gtrsim 600$~GeV. This limit weakens to
$m_{\stone}\gtrsim 450$~GeV for heavy neutralinos, similarly to the
case when the decay mode $\stone\to b\chaone$
opens~\cite{stop-atlas-papers,stop-atlas-conf,stop-cms-conf}.
Compressed spectra reduce the experimental reach even further, leading
to allowed top masses just above the top
mass~\cite{light_stop_para}. The latter can then be searched for in
leptonic decay channels~\cite{light_stop_lep}, hadronic decay
channels~\cite{light_stop_had}, or new associated production
channels~\cite{light_stop_prod}. Obviously, pair production of the
light(est) third generation squark should come with a large LHC cross
section. On the other hand, we will see that associated stop--chargino
production can give rise to sizeable event numbers and can therefore
play a key role in searches for neutralinos and charginos. Moreover,
through their off--shell effects the three production processes,
namely stop pair production, chargino pair production, and associated
stop--chargino production are closely tied at the next-to-leading
order (NLO) level.\bigskip

The challenge in LHC analyses assuming very specific supersymmetric
particle spectra is the availability of precision predictions.  For
example, the original theoretical studies of squark production at
hadron colliders usually assume degenerate squark
masses~\cite{squarkpairLO,squarkpairNLO,gluinopairNLO,squarkgluinoNLO,prospino}.
The NLO computation of stop pair production allows for separate stop
masses~\cite{stopNLO}.  More recently, improved predictions have
become available including electroweak~\cite{squarkEW}, higher-order
QCD~\cite{squarkgluinoNNLO} and resummed
corrections~\cite{squarkgluinoResummed} mostly assuming simplified
supersymmetric mass patterns. Efforts in matching fixed--order NLO
predictions to parton showers are presently
underway~\cite{Gavin:2013kga}.  In this paper we discuss the extension
of automized \textsc{MadGolem}
setup~\cite{Binoth:2011xi,GoncalvesNetto:2012yt,madgolem_nonsusy} to
third generation squark rates, including bottom densities, to
NLO accuracy\footnote{The \textsc{MadGolem} code
  with pre-compiled processes is available from the authors upon
  request.}. The key advantage of this approach is its complete
flexibility in the model assumptions which are defined in the
\textsc{Madgraph} framework. Correspondingly, in this paper we for the
first time provide precision predictions for fully exclusive event
rates and with completely flexible model parameters for the upcoming
LHC runs. Assuming for example a simplified model of top squarks and
charginos/neutralinos the processes discussed in this paper together
with chargino pair production will allow for a complete
NLO treatment of LHC rates and distributions with
\textsc{MadGolem}.\bigskip

The remainder of this paper is organized as follows: in
Section~\ref{sec:framework} we present the framework of our
calculation, describing both the automated \textsc{MadGolem} setup and
the MSSM benchmarks we use for our phenomenological analysis.  In
Section~\ref{sec:results} the total and differential NLO rates are
portrayed and analyzed for three representative processes: i) stop
pairs $pp\to\stop_1\stop^*_1$; ii) sbottom pairs $pp\to\sbot_1\sbot^*_1$; and
iii) associated stop--chargino  production $pp\to\stop_1\chaone$.  The
role of the incoming bottoms and their description within the
4-flavor and 5-flavor schemes is addressed in a dedicated
subsection.  We summarize our results in Section~\ref{sec:summary}.
Details on the renormalization scheme and analytical
expressions for the relevant ultraviolet (UV) counterterms and
renormalization constants are given in the Appendix.

%%%%%%%%%%%%%%%%%%%%%%%%%%%%%%%%%%%%%%%%%%%%%%%%%%%%%%%%%%%%%%%%%%%%%%%%
\section{Setup}
\label{sec:framework}

We compute the total and differential NLO rates, including all QCD and
SUSY-QCD corrections in the fully automated \textsc{MadGolem}
setup~\cite{Binoth:2011xi,GoncalvesNetto:2012yt,madgolem_nonsusy}.
\textsc{MadGolem} is an independent, modular add-on to
\textsc{MadGraph}~\cite{mg4,mg5}. It generates all tree--level
diagrams in the \textsc{MadGraph}~v4.5 framework and uses
\textsc{Qgraf}~\cite{qgraf} and \textsc{Golem}~\cite{golem,golem_lib}
for the one-loop amplitudes. The relevant counterterms,
Catani--Seymour dipoles~\cite{catani_seymour} and subtraction terms
for on--shell divergences~\cite{squarkgluinoNLO,on-shell} are part of
the automated setup.

The NLO effects split into real and virtual corrections. Real
corrections originate from radiation off the initial or final
state. Their infrared (IR) divergences are subtracted by means of
massive Catani--Seymour dipoles~\cite{catani_seymour}. In addition to
the SM dipoles \textsc{MadGolem} provides squark dipoles for
final--final and final--initial singularities as an extension to
\textsc{MadDipole}~\cite{maddipole}.  We retain the dependence on the
FKS-like cutoff $\alpha$~\cite{alpha} to include more ($\alpha\to 1$)
or less ($\alpha\ll 1$) of the finite phase space in the dipole
subtraction. Our default choice is $\alpha=1$.

Virtual corrections arise from gluons and gluinos. The standard 't
Hooft--Feynman gauge for internal gluons avoids higher rank loop
integrals. Dimensional regularization breaks supersymmetry via the
mismatch between the two fermionic gluino components and the
$(2-2\epsilon)$ degrees of freedom of the transverse gluon
field. \textsc{MadGolem} restores it using finite counter
terms~\cite{squarkgluinoNLO,Martin:1993yx}.  The renormalization of
the third generation squark sector is more complex than for the first
and second generations, due to massive quarks and the corresponding
squark mixing. Details on the renormalization procedure are specified
in the Appendix.

Finally, we need to remove divergences due to intermediate resonant
states. One example is the well known appearance of on--shell gluinos as part of the
NLO squark pair production via the three-body subprocess
$qg\to\sq\go\to\sq\sq^*q$~\cite{squarkgluinoNLO,on-shell}. The caveat
is twofold: on the one hand, these configurations lead to a phase
space divergence. On the other hand, they can give rise to double
counting once all squark and gluino NLO production rates are
combined. Following the \textsc{Prospino} scheme we remove the
on--shell divergences locally through a point--by--point subtraction
over phase space. Off--shell contributions in the zero--width limit
are genuine parts of the NLO real emission.  This procedure preserves
the gauge invariance as well as the spin correlations between the
intermediate particles and the final state. The subtraction terms in
the corresponding \textsc{MadOS} module have a Breit--Wigner shape and
are generated automatically. For further details we refer the reader
to Appendix~B of Ref.~\cite{GoncalvesNetto:2012yt}.

Throughout our calculation we use consistent LO and NLO parton
densities given by CTEQ6L1 and CTEQ6M with five active
flavors~\cite{cteq} for a 14~TeV LHC. For a dedicated analysis of the 4-flavor
scheme in Section~\ref{subsec:4vs5} we resort to the
NNPDF21\_FFN\_NF4\_100~\cite{Ball:2011mu}
parton densities.  The strong coupling constant $\alpha_s(\mu_R)$ is
evaluated using 2-loop running from $\Lambda_\text{QCD}$ to the
required renormalization scale $\mu_R$, again with five active
flavors. Unless stated otherwise, we use a common central value of the
average final state mass for the renormalization and factorization
scales, $\mu_R=\mu_F\equiv\mu^0=(m_1+m_2)/2$. From previous studies we
know this choice to yield perturbatively stable
results~\cite{squarkgluinoNLO}.\bigskip

So-called phenomenological MSSM (pMSSM)
scenarios~\cite{Berger:2008cq,Cahill-Rowley:2013gca} describe the
supersymmetric parameter space featuring: i) R-parity and
CP-conservation; ii) minimal flavor violation at the electroweak
scale; iii) degenerate light--flavor squarks; and iv) negligible
Yukawa couplings and trilinear interactions for the first and second
generations.  These requirements shrink the general MSSM parameter
space to 19 free parameters, while they do not demand any specific
assumption neither on the underlying high--scale model nor on the
SUSY-breaking pattern.

We survey representative pMSSM benchmarks which satisfy all
major experimental constraints from direct collider searches, flavor
physics and electroweak precision data. All of them predict a Higgs
boson mass of $m_h \simeq 126$~GeV and a LSP relic density close to
the observed $\Omega h^2=0.1126\pm 0.0036$.  Specifically, we consider
the following scenarios: Model 401479, 1889214, 2178683, 2342344, and
2948967, in the notation of Ref.~\cite{Cahill-Rowley:2013gca}.  Since
the stop and sbottom masses range around 1~TeV we find pair production
rates barely reaching $\mathcal{O}(1)$~fb at the topmost, with typical
quantum effects ranging $K=1.5-2$ for an LHC energy of 14~TeV. To
study light stop and sbottom production we consider more specific
scenarios:

\paragraph{Natural SUSY:}

Natural SUSY (NSUSY) features soft-breaking squark masses inversely
proportional to their Higgs couplings, combined with a small Higgsino
mass term~\cite{nsusy}. Typical mass spectra feature sub-TeV heavy--flavor squarks,
weak-scale higgsino-like neutralinos and charginos, and multi-TeV
gluinos.  We first examine two NSUSY scenarios~\cite{nsusy-benchmarks}
marked as `NSUSY1' and `NSUSY2' in Table~\ref{tab:spectra}.

\paragraph{Natural-like constrained MSSM:}

Complementary NSUSY benchmarks follow
Ref.~\cite{Buchmueller:2013exa}. We start from constrained MSSM
(CMSSM) points~\cite{sps} and assume non-universal squark
soft-breaking masses at the GUT scale, leading to light third
generation squarks, while the first and second generation stay above
1~TeV. The resulting CMSSM scenarios are labelled as `NS-CMSSM\#' in
Table~\ref{tab:spectra}.

\paragraph{Light SUSY:}

Finally, we consider configurations defined in
Ref.~\cite{Han:2013mga}. They are not linked to GUT-scale assumptions
but inspired by their implications on Higgs physics.  For all
scenarios we employ \textsc{SoftSUSY}~\cite{Allanach:2001kg} to
compute the physical particle masses.

%------------------------------------------
\begin{table}[t]
\begin{center}
\begin{tabular}{|ll||rr|rr|rr|r|rr|} \hline
 && $m_{\stop_1}$ & $m_{\stop_2}$ & $m_{\sbot_1}$ & $m_{\sbot_2}$ &
    $m_{\tilde u_{L/R}}$ & $m_{\tilde d_{L/R}}$ &  $m_{\tilde{g}}$ &
    $m_{\chaone}$ & $m_{\neutone}$ \\ \hline 
 a. \quad &NSUSY 1 & 434.9 & 990.3 & 891.6 & 1356.9 & 
          13453/13444 & 13453/13444 & 3202.6 & 222.6 & 216.8  \\ 
 &NSUSY 2 & 874.4 & 1614.4 & 1580.8 & 1969.9 & 
          7389.6/7381.1 & 7389.8/7361.1 & 2770.0 & 151.9 & 146.0 \\
\hline
 b. &NS-CMSSM 10.2.2 & 398.4 & 682.5 & 572.4 & 684.6 & 
       5075.1/5071.7 & 5075.6/5071.7 & 1354.7 & 425.4 & 231.3 \\
 &NS-CMSSM 40.2.2 & 255.8 & 611.8 & 471.5 & 649.8 & 
       5055.8/5054.2 & 5056.2/5054.1 & 1251.8 & 398.7 & 211.9 \\
 &NS-CMSSM 40.3.2 & 320.0 & 610.9 & 528.3 & 771.7 & 
       5015.5/5017.6 & 5015.9/5017.9 & 974.6 & 288.6 & 157.1  \\
\hline
 c. &Light SUSY & 374.4 & 2022.9 & 387.9 & 2011.6 & 
       2013.2/2011.8 & 2014.6/2011.6 & 1102.3 & 498.9 & 301.3 \\ 
\hline
\end{tabular}
\end{center}
\caption{Sparticle masses (in GeV) for the different MSSM benchmarks
 employed in our calculation.}
\label{tab:spectra}
\end{table}
%------------------------------------------

%%%%%%%%%%%%%%%%%%%%%%%%%%%%%%%%%%%%%%%%%%%%%%%%%%%%%%%%%%%%%%%%%%%%%%%%
\section{Results}
\label{sec:results}

In the stop--chargino sector the pair production of stop pairs and of
a stop in association with a chargino are tied together, because at
the NLO their on--shell and off--shell contributions
overlap. We start by largely reviewing stop (and sbottom) pair
production with a focus on light states and on their kinematic
distributions. Then we move on the associated production process,
which given the overwhelming backgrounds to chargino pair production
might well be the most promising signature for charginos at the LHC.

%%%%%%%%%%%%%%%%%%%%%%%%%%%%%%%%%%%%%%%%%%%%%%%%%%%%%%%%%%%%%%%%%%%%%%%%
\subsection{Stop and sbottom pair production}
\label{subsec:squarkpair}

\begin{figure}[b!]
\begin{center}
 \includegraphics[height=0.06\textheight]{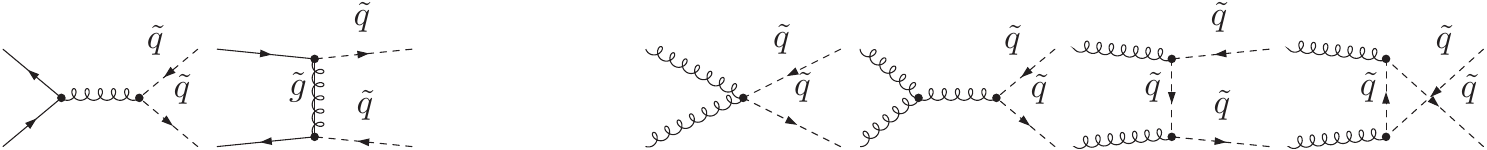} \\[4mm]
 \includegraphics[height=0.06\textheight]{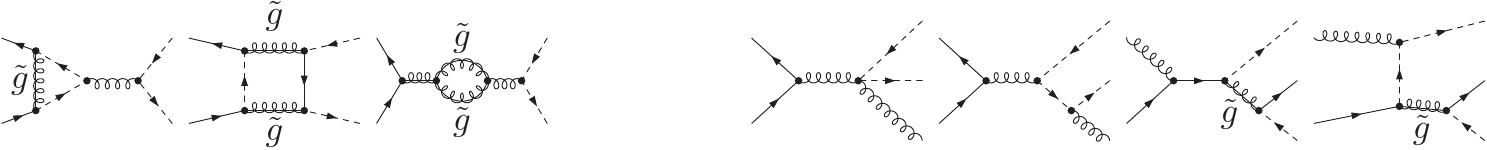}
\end{center}
\caption{Feynman diagrams for squark--antisquark production at LO and
  at NLO. In the upper row we separately show quark and gluon initial
  states. The gluino exchange only features for sbottom pairs.  For
  NLO we display representative vertex, box, and self-energy topologies as
  well as real corrections.}
\label{fig:feyn-stoppair}
\end{figure}

We first compute the cross sections 
for the pair production of heavy--flavor squarks:  
\begin{alignat}{5}
 pp\to\stop_1\stop^*_1,\;\sbot_1\sbot^*_1 \; .
 \label{eq:squarkpair}
\end{alignat}
The global $K$ factor is defined as
$K\equiv\sigma^\text{NLO}/\sigma^\text{LO}$. Some corresponding
Feynman diagrams are displayed in Figure~\ref{fig:feyn-stoppair}.  At
variance with the first and second generation squark production, the
flavor--locked $q\bar q$-initiated diagram with a gluino exchange is
absent for stop pairs. For sbottom pairs, this diagram
arises from the initial--state bottom quarks, which we include
consistently in the five flavor scheme (5FS). We evaluate all rates for an LHC energy of
14~TeV. Representative MSSM parameter space regions are covered by the benchmark
points defined in Section~\ref{sec:framework}. Unlike in
\textsc{Prospino} the general setup of \textsc{MadGolem} enables us to
separate all squark flavor and chirality without making assumptions on
the mass spectrum.\bigskip

%------------------------------------------
%------------------------------------------

Table~\ref{tab:squarkpair14} shows cross sections ranging from 10~fb
to $10^4$~fb for stop and sbottom masses up to the TeV range.  The
relative NLO correction is comparably stable at $K\sim1.5-1.9$. This
reflects the fact that the dominant NLO quantum effects arise from
QCD, while SUSY-QCD corrections are subleading. The very slightly
larger $K$ factors for sbottom pair production compared to their stop
counterparts are related to the additional one-loop effects from the
gluino $t$-channel exchange between the incoming bottom partons. 
It is also illustrative to compare the rates to their
light--flavor counterparts given in Table~2 of
Ref.~\cite{GoncalvesNetto:2012yt}. Novel one-loop contributions for
example from non-diagonal $\tilde{q}_1-\tilde{q}_2$ mixing
self-energies are intrinsically tied to the heavy-flavor squark
structure and account for the larger $K$-factors.\bigskip

%------------------------------------------
\begin{table}[t]
\begin{center}
\begin{tabular}{|ll||rrrr|rrrr|} \hline
 && \multicolumn{4}{c|}{$pp \to \stop_1\stop^*_1$}
  & \multicolumn{4}{c|}{$pp \to \sbot_1\sbot^*_1$}\\
 && $m_{\stop_1}$ [GeV] & $\sigma^{\text{LO}}$ [fb] & $\sigma^{\text{NLO}}$ [fb] & $K$ 
  & $m_{\sbot_1}$ [GeV] & $\sigma^{\text{LO}}$ [fb] & $\sigma^{\text{NLO}}$ [fb] & $K$ \\
\hline
a. \quad &NSUSY 1 & 434.9 & 881  & 1380 & 1.57 
                  & 891.6 & 10.6 & 18.0 & 1.70 \\
&NSUSY 2 & 874.4  & 12.1 & 20.4 & 1.69 
         & 1580.8 & 0.11 & 0.23 & 1.87 \\
\hline
b. &NS-CMSSM 10.2.2 & 398.4 & 1430  & 2210  & 1.54 
                    & 572.4 & 180   & 290   & 1.61 \\
   &NS-CMSSM 40.2.2 & 255.8 & 14800 & 21800 & 1.47 
                    & 471.5 & 558   & 882   & 1.58 \\      
&NS-CMSSM 40.3.2 & 320.0 & 4680  & 7010  & 1.50 
                 & 528.3 & 28900 & 46200 & 1.60 \\ 
\hline
c. &Light SUSY & 374.4 & 2010 & 3080 & 1.53 
               & 387.9 & 1660 & 2550 & 1.53 \\ 
\hline
\end{tabular}
\end{center}
\caption{Total rates and corresponding $K$ factors for third generation
 squark pair production at the 14~TeV LHC.}
\label{tab:squarkpair14}
\end{table}
%------------------------------------------

%------------------------------------------
\begin{figure}[b!]
 \includegraphics[height=0.235\textheight]{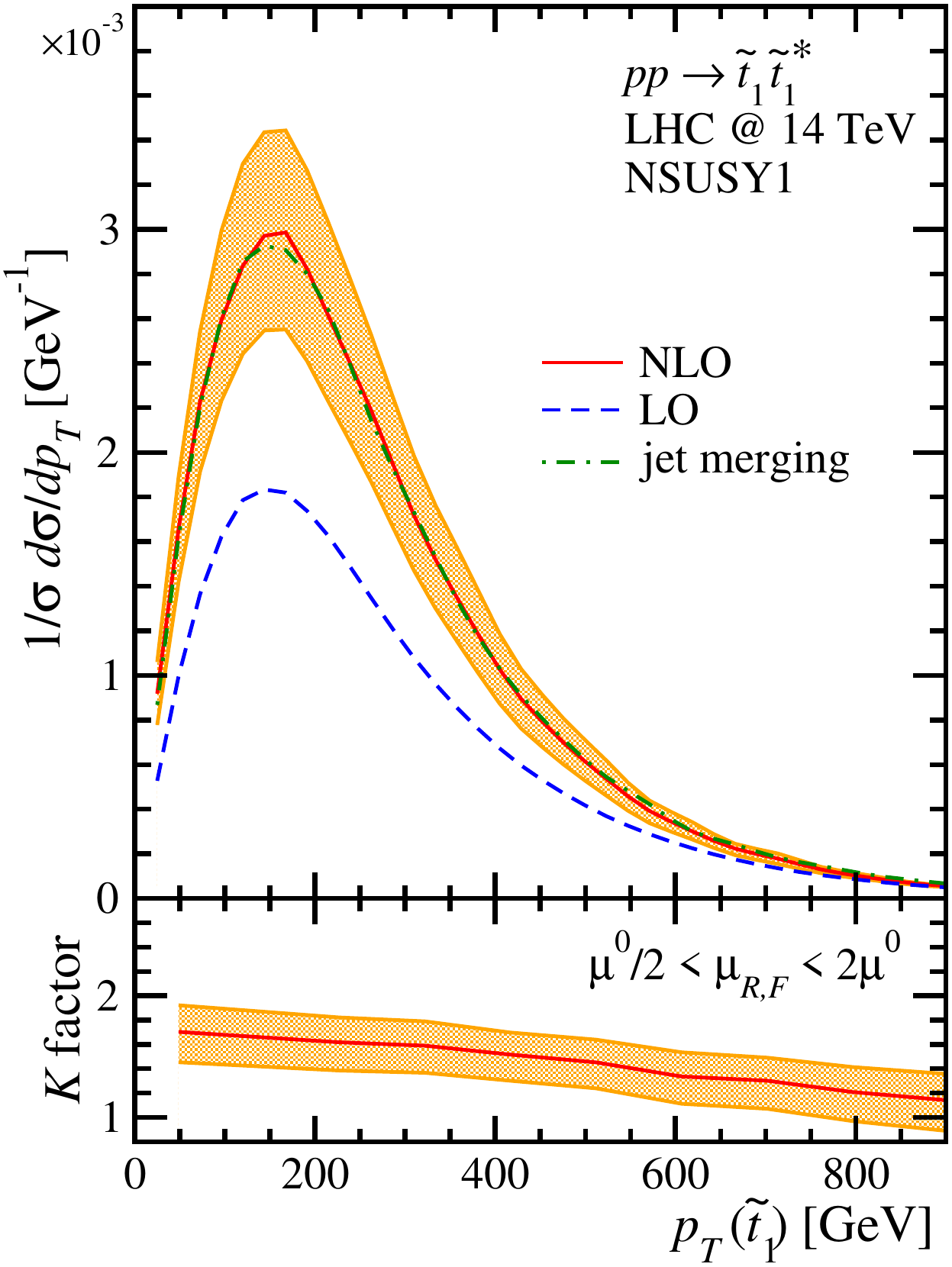} 
 \includegraphics[height=0.235\textheight]{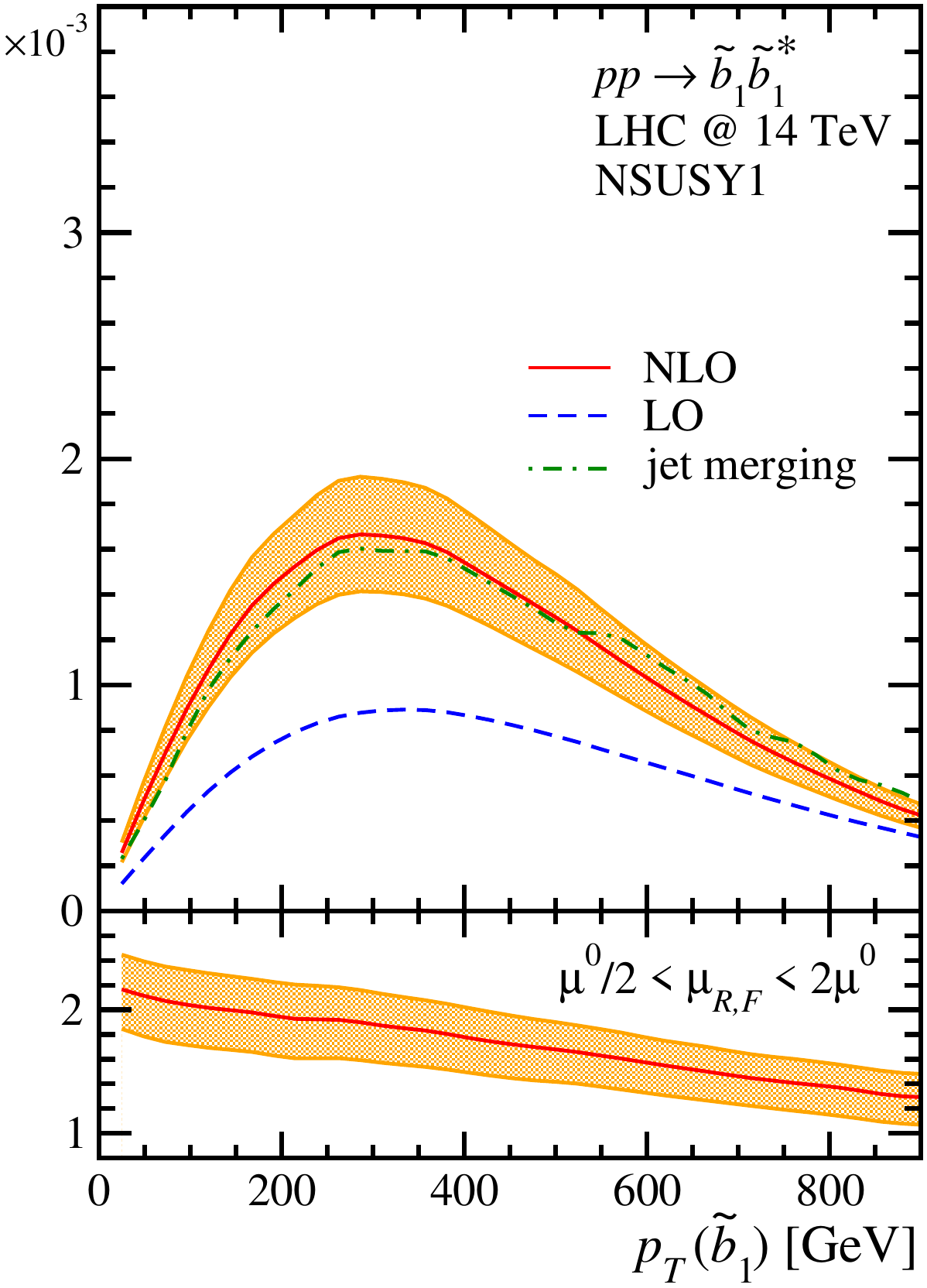}\quad
 \includegraphics[height=0.235\textheight]{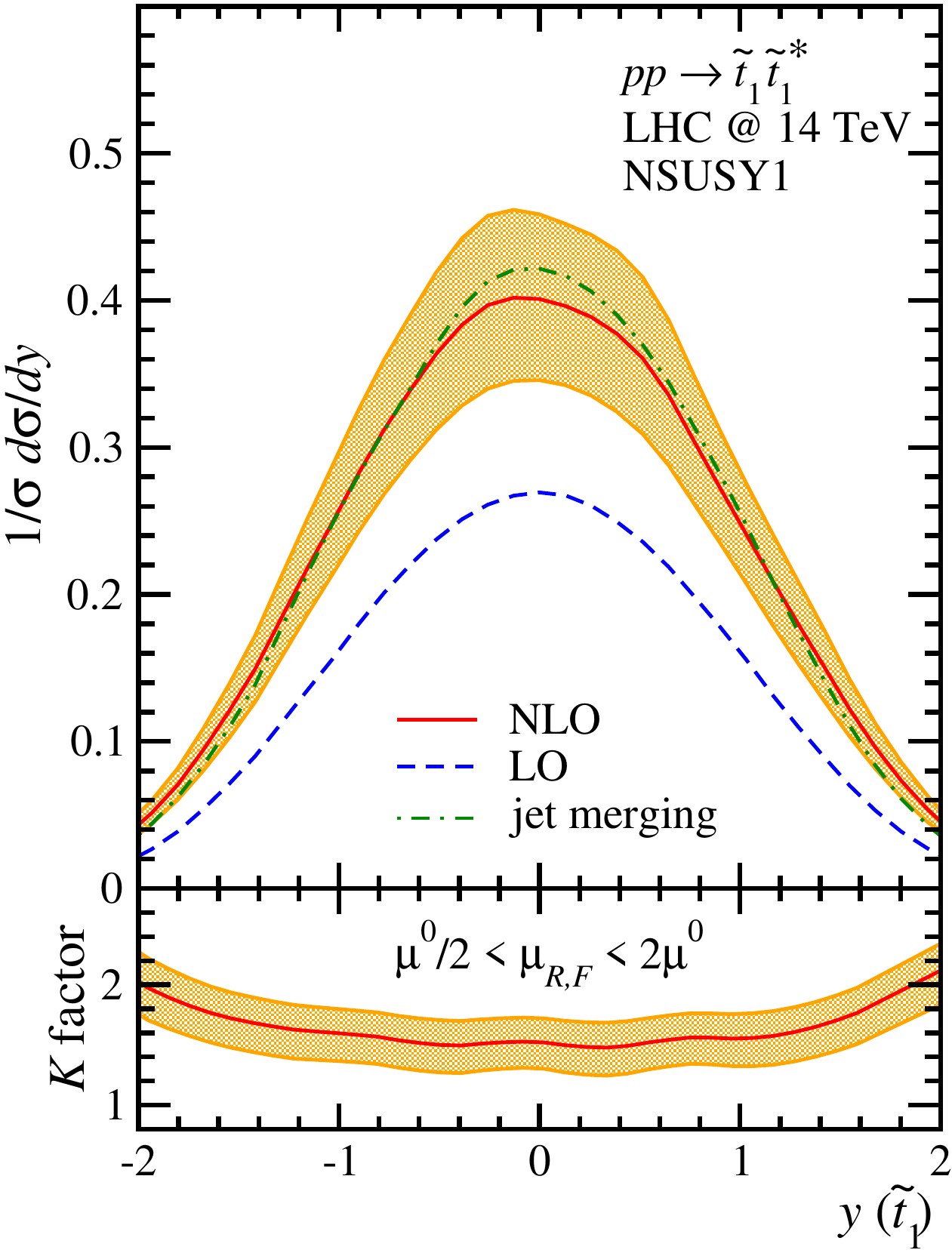} 
 \includegraphics[height=0.235\textheight]{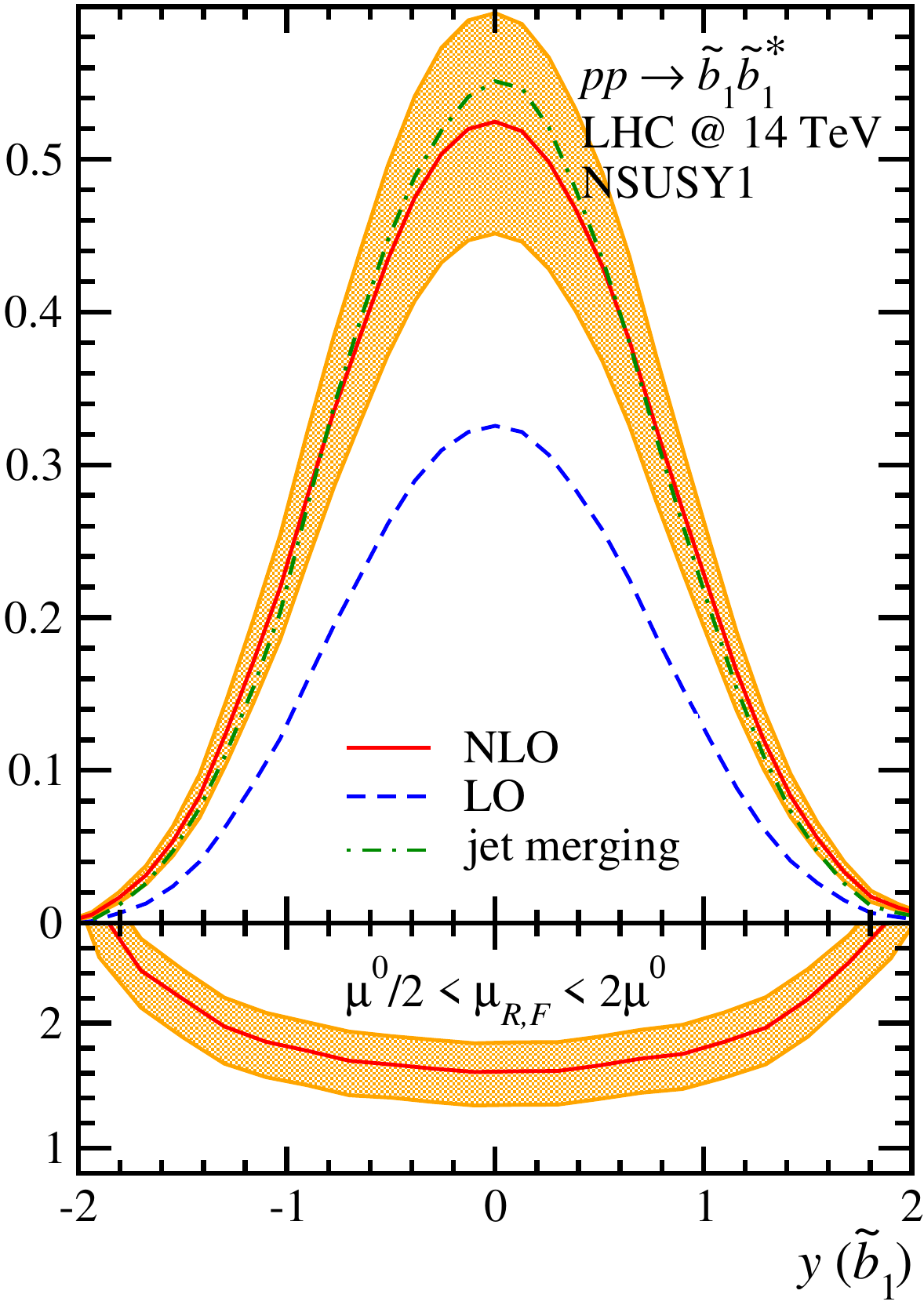}
\caption{Normalized fixed--order NLO and tree--level merged
  distributions for $pp\to\stop_1\stop^*_1/\sbot_1\sbot^*_1$ as a function of
  the transverse momentum and rapidity.  All MSSM
  parameters are fixed to the NSUSY1 benchmark in
  Table~\ref{tab:spectra}.  The error band for the NLO distributions
  corresponds to a variation $\mu^0/2<\mu_{R,F}<2\mu^0$.  The lower
  panels show the corresponding $K$ factors.}
\label{fig:pair}
\end{figure}
%------------------------------------------

The automated \textsc{MadGolem} setup also provides predictions beyond
total rates through the output of weighted event samples for the
regularized virtual and real correction channels.  They can be used to
plot any NLO distribution which is consistently defined in
perturbative QCD.  In Figure~\ref{fig:pair} we display the
distributions of the stop and sbottom transverse momentum and
rapidity, including their scale uncertainties. As a representative
MSSM configuration we choose NSUSY1 as defined in
Table~\ref{tab:spectra}.  The distributions are normalized to unity
for the central scale choice $\mu^0$; the uncertainty envelope is
generated through a parallel change of the renormalization and
factorization scales in the conventional range $\mu^0/2-2\mu^0$.  The
transverse momentum of the stops has a strong peak slightly below
200~GeV.  The fact that the lightest sbottom in this scenario is
roughly twice as heavy as the stop explains why the sbottom distributions
show a significantly harder and more central profile.

The stabilization of the (unphysical) scale dependence, 
which shrinks from $\mathcal{O}(60\%)$ at LO down to 
$\mathcal{O}(30\%)$ at NLO, reflects the improved
precision of the higher order computation. The lower panels in
Figure~\ref{fig:pair} show the corresponding $K$ factors, also with
their scale uncertainty.  Its sizeable variation shows that the
higher--order corrections cannot simply be described by a constant $K$
factor. While in the past the LHC analyses would typically rely on
central threshold production of heavy particles, more targeted
analyses for example for light stops will be sensitive to specific
phase space regions. In the case of stop pairs the $K$ factor at
threshold ranges around $1.7 \pm 0.3$; transverse stop momenta in the
300~GeV range still leave a sizeable fraction of events, but with a
$K$ factor around 1.6. While this effect is not dramatic given the
quoted NLO error bars, it should be considered once
we include higher order corrections.\bigskip

In addition to the fixed--order NLO results in Figure~\ref{fig:pair} we also
show the tree--level multi--jet merging results.  For the latter we
include up to two additional hard jets combined with the
\textsc{Pythia}~\cite{pythia} parton shower and merge them via the MLM
scheme~\cite{mlm} in \textsc{MadGraph5}~\cite{madgraph_merging}.  We
consider three setups: i) up to one hard (gluon) jet; ii) up to two
hard (gluon) jets; iii) up to one hard (light quark and gluon)
jet. The first two approaches avoid any potential issues with
on--shell singularities, which means they are only sensitive to QCD
radiation. The third case involves a subset of bottom--initiated
contributions which are sensitive to an intermediate on--shell gluino,
just like the case shown in Figure~\ref{fig:feyn-stoppair}. To improve
speed we regularize this divergence by means of a numerical recipe
implemented in \textsc{MadGraph}~\cite{mg5}, which subtracts all
events close to on--shell poles.  While this subtraction method is not
equivalent to the rigorous \textsc{Prospino} scheme and does not
provide a well--defined zero--width limit, we have checked that it
gives numerically similar results, as long as we consider normalized
distributions\footnote{For predictions of general distributions
  including the proper normalization only the \textsc{Prospino} scheme
  for on--shell subtraction will be consistent and numerically reliable.}.

This way we find that all three--jet merging setups agree very well
with each other and with the fixed--order NLO result. Neither
including a second hard gluon jet nor a light quark hard jet leads to
any significant difference compared to the emission of up to one
gluon. The observed shift towards slightly harder and more central
squarks in the merged sample can be attributed to the additional
recoil jets. All this is a common feature in the production of heavy
colored resonances, which is characterized by a single hard scale well
above the typical jet momenta required by inclusive
searches~\cite{qcd_radiation,madgraph_merging}. The same behavior can
be seen in light--flavor squark and gluino
production~\cite{GoncalvesNetto:2012yt}.  Once the double counting
from the on--shell states is removed, the NLO real emission almost
entirely consists of soft and collinear gluons, properly accounted for
by the parton shower. As a note of caution we nonetheless keep in mind
that once an experimental analysis becomes sensitive to the jet recoil
the simulation should include matrix element and parton shower
merging~\cite{autofocus}. Examples for that are mono-jets
searches as well as the analysis of forward (tagging) jets in the
top/stop pair sample.

%%%%%%%%%%%%%%%%%%%%%%%%%%%%%%%%%%%%%%%%%%%%%%%%%%%%%%%%%%%%%%%%%%%%%%%%
\subsection{Associated stop--chargino production}
\label{subsec:stopcha}

Next, we consider top squark production in association with
the lightest chargino, 
\begin{alignat}{5}
 pp\to\stone\tilde\chi^-_1 \; .
\label{eq:stopcha}
\end{alignat}
The LO process is flavor--locked, meaning that at parton level it
relies on an incoming bottom quark.  Throughout our calculation we
resort to a five flavor scheme and use consistent parton densities and
their corresponding $\alpha_s$ values.  One technical caveat is that a
massless bottom quark is incompatible with the left--right mixing in
the sbottom sector.  We circumvent this problem by assuming massless
external bottom lines and propagators, while keeping a finite Yukawa
coupling.  The renormalization of the (massless) bottom quark and the
(massive) bottom Yukawa is described in the Appendix.  Further
considerations on the flavor scheme choice are addressed in detail in
subsection~\ref{subsec:4vs5} below.\bigskip

%------------------------------------------
\begin{figure}[b!]
\begin{center}
 \includegraphics[height=0.055\textheight]{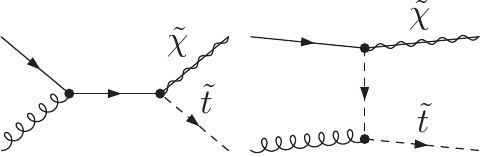} \\[4mm]
 \includegraphics[height=0.07\textheight]{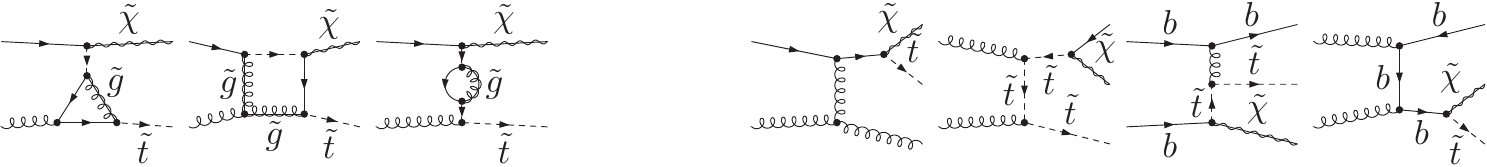}
\end{center}
\caption{Feynman diagrams for the production of a top squark
  associated with a chargino at LO and at NLO. For the NLO case we
  show representative vertex, box, and self-energy topologies, as well
  as real corrections.}
\label{fig:feyn-stch}
\end{figure}
%------------------------------------------

As usual, NLO corrections proceed via virtual QCD and SUSY-QCD
contributions and real emission.  The relevant Feynman diagrams are
illustrated in Figure~\ref{fig:feyn-stch}. When we include bottom
densities the perturbative counting requires some care, in particular
for the genuine next-to-leading contributions which are no more
flavor--locked. For example, off--shell stop pair production in
quark--antiquark scattering or gluon fusion with a subsequent stop
decay into a chargino contributes to the NLO
rate. Strictly speaking, these diagrams are suppressed by one power of
$\alpha_{EW}$ compared to the associate production at LO in the 5FS. On
the other hand, modulo large logarithms the bottom density can be
considered subleading to the gluon density by one power of $\alpha_s$,
in particular when the resummed logarithms are only moderately
large. This means that we have to rely on other effects, like the size
of off--shell vs on--shell contribution to ensure the perturbative
stability of the rate predictions. Obviously, these complications will
not be reflected in the scale variation as a measure of the
theoretical uncertainty.

In addition, the real emission includes gluon splitting into a pair of
bottoms in the initial state; this is the same diagram which defines
the bottom parton densities in the collinear limit. However, as for
any initial state, this collinear divergence is absorbed in the parton
densities. In that sense the NLO computation in the 5FS can be viewed
as the leading combination of the 4-flavor and the 5-flavor
predictions~\cite{4vs5higgs}. The relative size of the two
contributions can be adjusted by varying the bottom factorization
scale, as described in the following section.  As discussed in the
next section we use the average heavy mass of the final state as
central renormalization and factorization scale.\bigskip

%------------------------------------------
\begin{table}[t]
\begin{center}
\begin{tabular}{|ll||rrrrr|} \hline
 && \multicolumn{5}{c|}{$pp \to \stop_1\tilde{\chi}_1^-$}\\
 && $m_{\stop_1}$ [GeV] & $m_{\chaone}$ [GeV] & $\sigma^\text{LO}$ [fb] & $\sigma^\text{NLO}$ [fb] & $K$  \\ \hline
a. \quad &NSUSY1 & 434.9 & 222.6 & 40.97 & 49.98 & 1.22   \\
         &NSUSY2 & 874.4 & 151.9 &  1.94 & 2.51 & 1.29    \\ 
\hline
b. &NSCMSSM-10.2.2 & 398.4 & 425.4 & 13.40   & 20.14  & 1.50   \\
   &NSCMSSM-40.2.2 & 255.8 & 398.7 & 47.83 & 71.21 & 1.48  \\      
   &NSCMSSM-40.3.2 & 320.0 & 288.6 & 53.39 & 78.94 & 1.48  \\ \hline
c. &Light SUSY     & 374.4 & 498.9 & 9.96 & 10.51 & 1.05   \\ \hline 
\end{tabular}
\end{center}
\caption{Total rates and corresponding $K$ factors for single stop
  production associated with the lightest chargino at the 14~TeV LHC.}
\label{tab:stopchargino}
\end{table}
%------------------------------------------

We first use \textsc{MadGolem} to compute the total rates for
stop--chargino production for the representative MSSM benchmarks
defined in Table~\ref{tab:spectra}. The results are documented in
Table~\ref{tab:stopchargino}.  The rates can be as large as 80~fb and
show tempered quantum effects $K\sim 1.1-1.5$, typically below their
first-generation squark counterparts~\cite{Binoth:2011xi}.  The main
reason for the moderate quantum corrections are the smaller bottom
luminosities, which suppress flavor--locked contributions $pp(bb) \to
\stone\tilde\chi^-_1\,b$ (cf. second diagram from the right in
Figure~\ref{fig:feyn-stch}) as compared to \eg
$pp(dd)\to\tilde{u}_L\,\tilde\chi^-_1\,d$. The different $K$-factors can be 
basically attributed to the relative sizes of the couplings $\tilde{t}_{1,2}$-$\chaone$-$b$
in each scenario.  \bigskip

\begin{figure}[b!]
\includegraphics[width=0.8\textwidth]{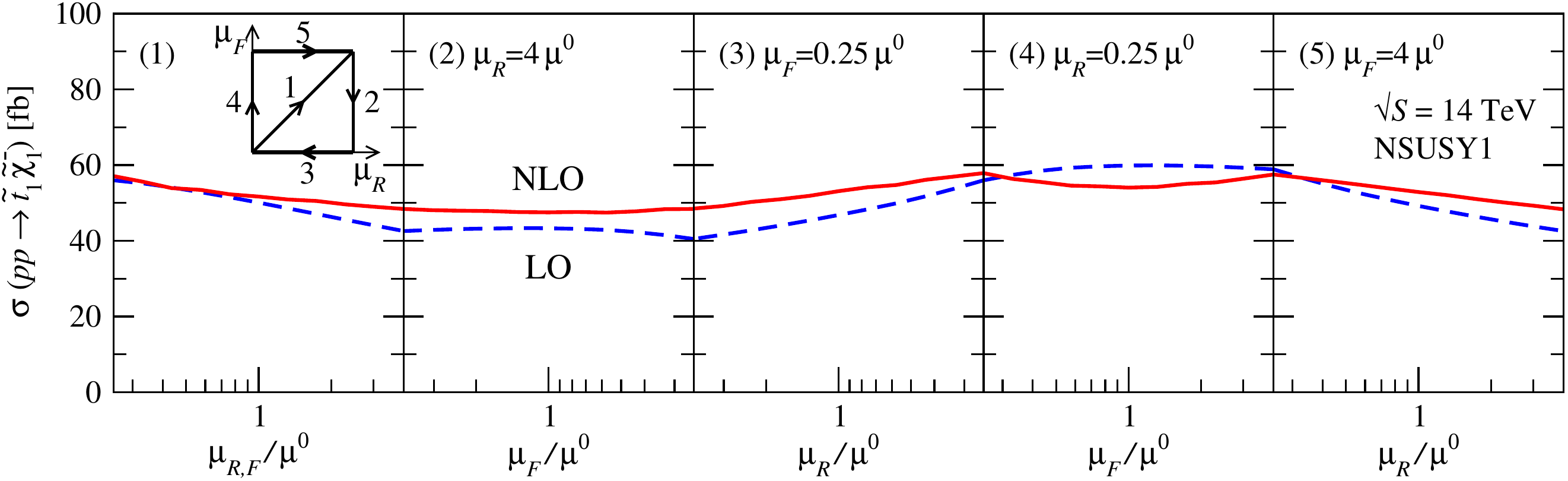}
\caption{Fixed--order rates for stop production associated with the
  lightest chargino. We show the scale variation in the $\mu_R-\mu_F$
  plane around the central scale $\mu^0 = (m_{\stone} +
  m_{\chaone})/2$.}
\label{fig:st1x1-scale}
\end{figure}

\begin{figure}[t]
\begin{center}
 \includegraphics[height=0.235\textheight]{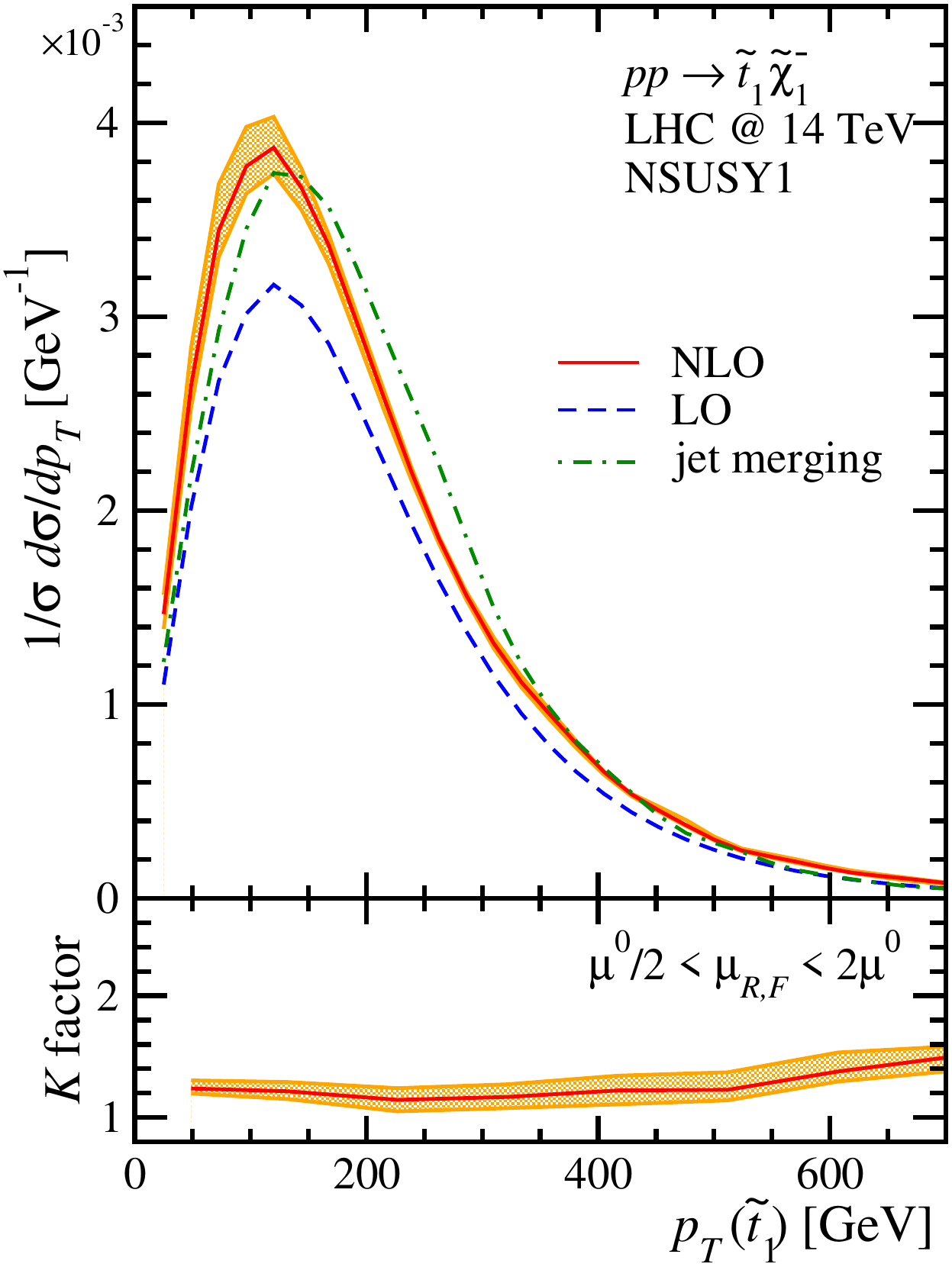} \qquad
 \includegraphics[height=0.235\textheight]{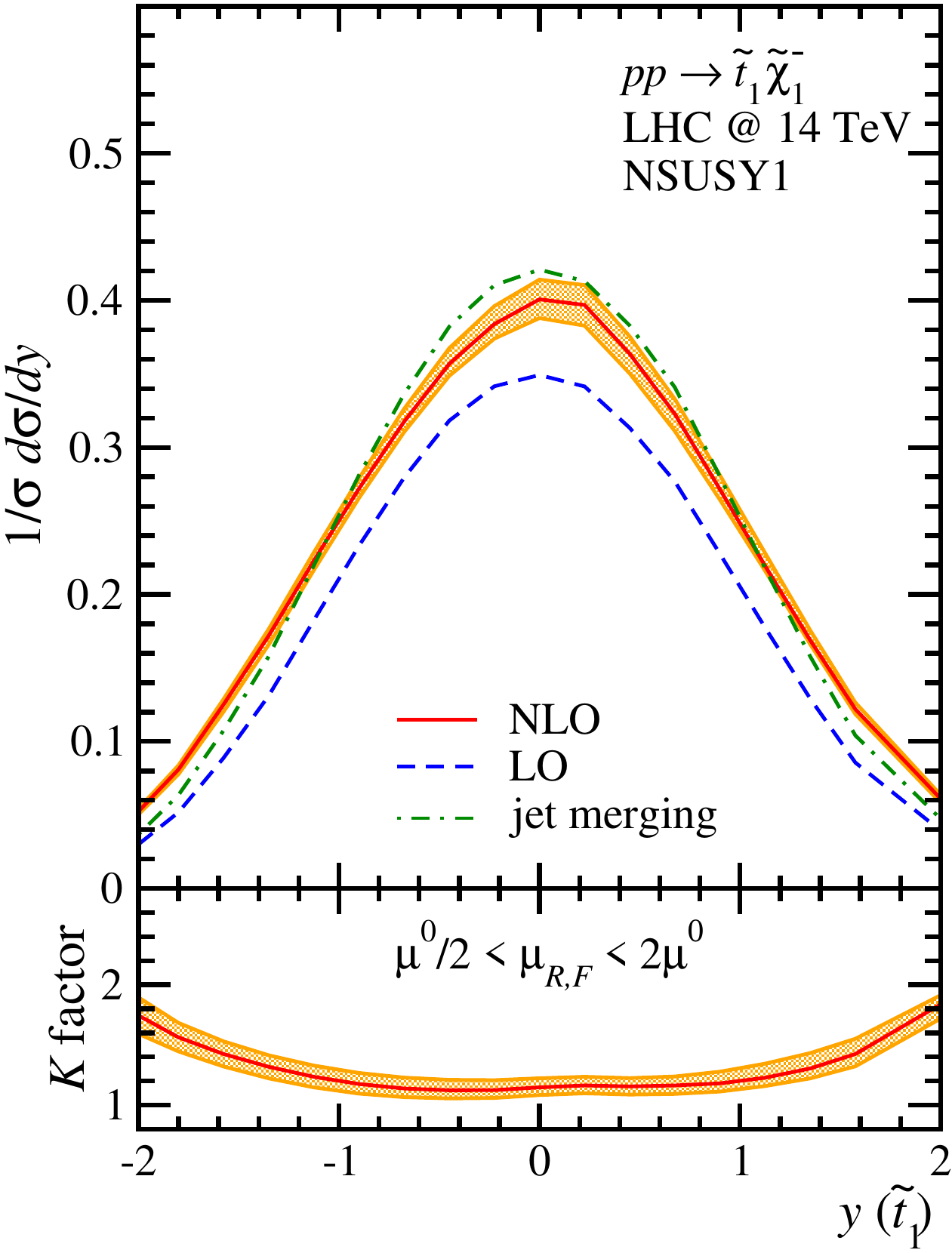}
\end{center}
\caption{Normalized fixed--order NLO and tree--level merged
  distributions for $pp\to\stop_1\tilde\chi_1^-$ as a function of
  the stop transverse momentum and rapidity.  All MSSM
  parameters are fixed to the NSUSY1 benchmark in
  Table~\ref{tab:spectra}.  The error band for the NLO distributions
  corresponds to a variation $\mu^0/2<\mu_{R,F}<2\mu^0$.  The lower
  panels show the corresponding $K$ factors.}
\label{fig:st1x1}
\end{figure}

\begin{figure}[t]
\begin{center}
 \includegraphics[height=0.2\textheight]{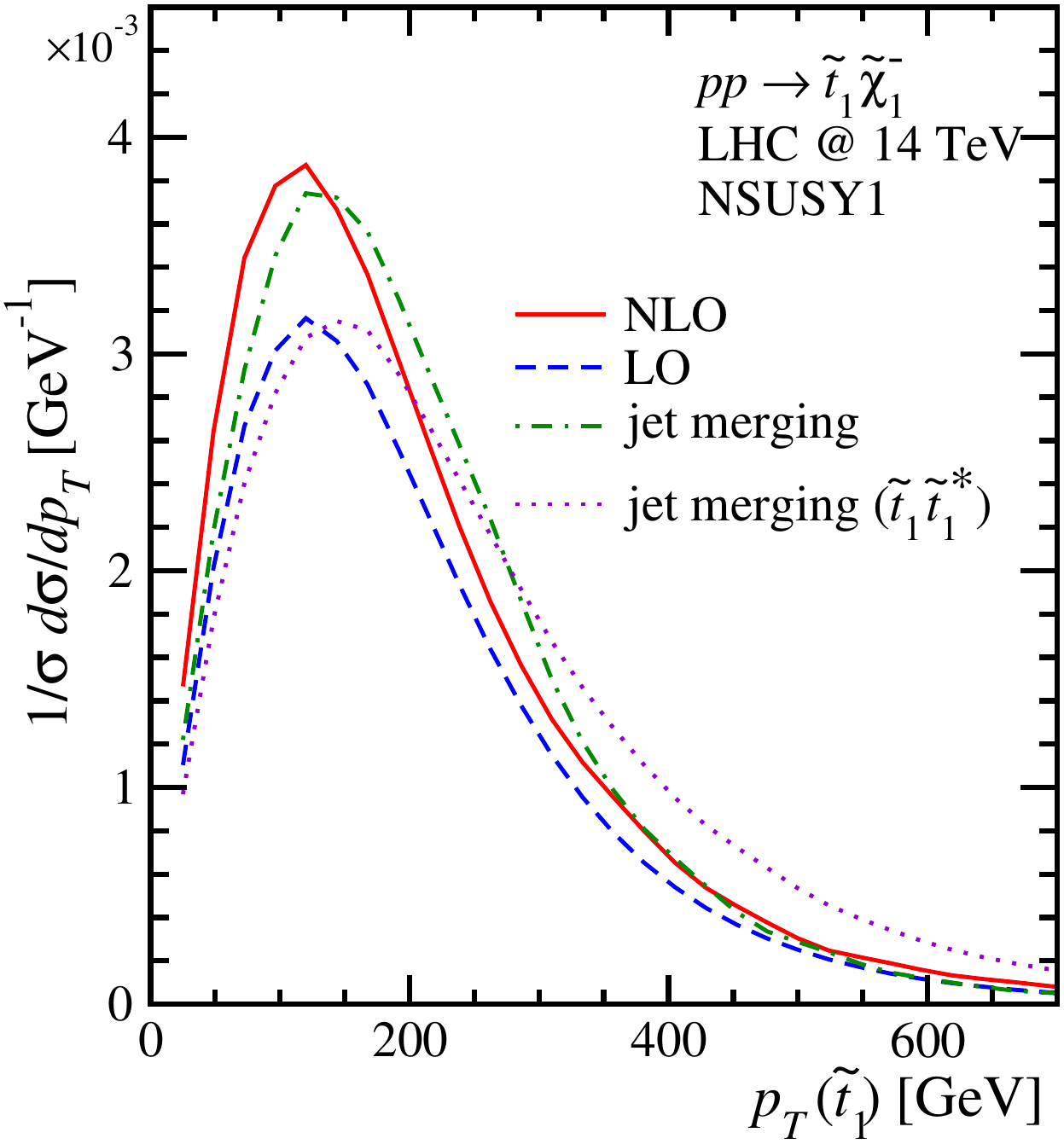} \qquad
 \includegraphics[height=0.2\textheight]{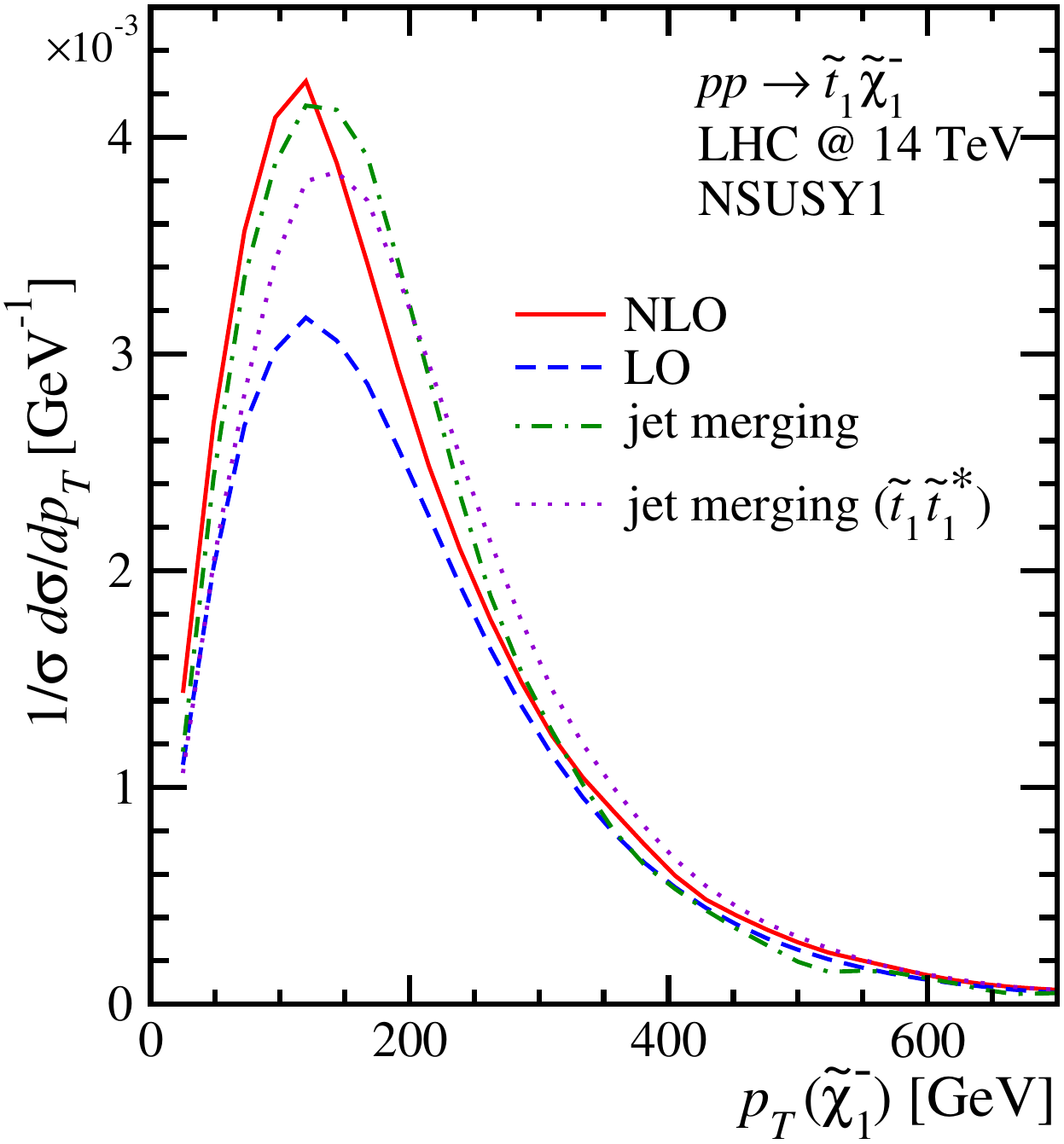} 
\end{center}
\caption{Normalized distributions for
 $pp\to\stone\tilde\chi^-_1$ as a function of the 
  stop and chargino transverse momentum. We compare the tree-level 
  merged distributions retaining (dotted lines) and subtracting (dotted-dashed 
  lines) stop pair production.}  
\label{fig:st1x1_extra}
\end{figure}

In Figure~\ref{fig:st1x1-scale} we display the change of the total
stop--chargino rates under correlated and independent variations of
the renormalization and factorization scales.  These scales are varied
in the range $\mu^{0}/4 < \mu_{R,F} < 4\mu^{0}$ and follow the path
illustrate in the little square of the first panel.  In spite of its
semi--weak nature, $\sigma^{LO} \propto \alpha_{EW} \alpha_s$, the scale
dependence in this process is dominated by the renormalization scale.  The
fairly flat slope around the central scale, which for NSUSY1 reads $\mu^0 =
(m_{\stone} + m_{\chaone})/2 \simeq 328$ GeV, can be traced back to
the individual scale dependence of the different parton densities.
The valence quark, gluon and bottom densities largely balance each
other and lead to an accidentally small factorization scale
uncertainty. The renormalization scale variation corresponding to one
power of the strong coupling does not reflect any of the complications
associated with the perturbative series of this process, as discussed
above. This means that we cannot expect the usual scale variation in
the range $\mu^0/2 - 2\mu^0$ to provide a reliable estimate of the
theoretical uncertainty. Correspondingly, the improvement of the scale
dependence when including the NLO corrections in
Figure~\ref{fig:st1x1-scale} is rather limited.\bigskip

Going beyond the total rates we show the differential distributions in
the stop transverse momentum and rapidity, together with the
multi--jet merging results in the 5FS in Figure~\ref{fig:st1x1}.  We
find excellent agreement between the fixed--order and the jet merging
calculations, the latter featuring as usual slightly harder, more
central stops. As for the total rates, some of the key aspect of the
NLO corrections are not governed by the scale dependence, which means
that unlike for stop pair production the predictions from jet merging
are not covered by the NLO scale dependence.

In Figure~\ref{fig:st1x1_extra} we show in addition the
transverse momentum distributions for the stop and for the chargino
without subtracting stop pair production (denoted as $\stone\stone^*$). 
For stop pair production
the kinematics of the chargino are determined by the on--shell stop
decays, and it is by accident that the transverse momentum
distribution agrees with the associated production.  In contrast, the
transverse momentum of the stop should now be identical to
Figure~\ref{fig:pair}. While in associated production channel the
transverse momentum of the stop peaks at low values slightly above
100~GeV, for stop pairs the peak increases by almost 200~GeV. The
reason for this shift is that the incoming bottom--gluon system is
significantly softer than a mix of incoming gluon--gluon or
quark--antiquark pairs. When combining the two production channels at
NLO the \textsc{Prospino} scheme for the on--shell
subtraction allows for simply adding the two event samples correctly
keeping track of these distinct features.

%%%%%%%%%%%%%%%%%%%%%%%%%%%%%%%%%%%%%%%%%%%%%%%%%%%%%%%%%%%%%%%%%%%%%%%%
\subsection{4-flavor versus 5-flavor schemes}
\label{subsec:4vs5}

Owing to its flavor--locked nature, the associated production of
heavy--flavor squarks relies on initial-state bottoms. This is exactly
the same situation as for the associated production of a top quark
with a charged Higgs boson~\cite{4vs5higgs}.  Depending on the
relevant scales and observables of interest such processes can be
described either in a so-called 4-flavor (4FS) or a 5-flavor scheme
(5FS)~\cite{acot,4vs5higgs,Maltoni:2012pa}.\bigskip

In the 4FS bottom quarks are not considered partons inside
the proton, but they are generated dynamically via gluon splitting as
part of the hard process. The bottom mass acts as an infrared regulator
for inclusive observables, which means that the numerical predictions
are reliable as long as $m_b$ is not too different from the relevant
scales of the process. One advantage of this scheme is that it retains
the full bottom--mass dependence in the final state kinematics.\bigskip

The production of massive states with comparably low transverse
momenta in the gluon splitting will be dominated by collinear gluon
splitting. The bottom mass still acts as a mathematical cutoff, but it
also generates a large physical logarithm, so the final--state bottom
jets peak at large rapidity, $y_b\simeq 2$, and low transverse
momenta, $p_{T,b}\simeq m_b$. 
In this kinematic
regime the $b$-tagging becomes a challenge, so the experimentally
relevant observables do not include the bottom radiation. To predict
the bottom--inclusive rate we need to integrate over the phase space
of the bottom radiation, approaching the asymptotic form
\begin{alignat}{5}
 \frac{d\sigma[\stone\tilde\chi^-_1\bar b]}{dp_{T,b}} 
 \sim \cfrac{p_{T,b}}{p^2_{T,b} + m_b^2} 
 \qquad \text{and hence} \qquad
 \sigma[\stone\tilde\chi^-_1\bar b] 
 \sim
 \log \frac{p_{T,b}^\text{max}}{m_b} \; .
 \label{eq:asymptotic}
\end{alignat}
This potentially large logarithm is the same logarithm which appears
in any QCD splitting regularized by a finite quark
mass~\cite{acot}. Its resummation can be linked to introducing a
scale--dependent bottom parton density for the inclusive process
$gb\to\stone\tilde\chi^-_1$, defining the 5FS. 
This link identifies $p_{T,b}^\text{max}$ with the
factorization scale of the bottom parton, implying that such scale 
is not a free parameter, but
can be computed from the kinematics of the splitting process which
generates the bottom parton density.  Usually, in this approach the
now incoming bottom parton is assumed to be massless, but mass
corrections can be included in the parton splittings~\cite{acot}.

The 5FS is hence justified as long as we do not
reconstruct the final-state bottom jet and as long as the bottom mass
is negligible compared to the hard scale of the process.  While it is
easy to show that the factorization scale is linear with the hard scale,
$\mu_{F,b} \propto m_1 + m_2$, the associated prefactor is process
dependent. For example, the bottom factorization scale for a leading
$bq$ initial state should be larger than for a $bg$ initial state,
because incoming gluons carry less momentum to generate hard radiation
than incoming quarks.  The link between the improved perturbative
prediction and the appropriate scale choice has been thoroughly
studied in the context of Higgs production~\cite{4vs5higgs,bbhiggs}.\bigskip

%------------------------------------------
\begin{figure}
\begin{center}
 \includegraphics[width=0.3\textwidth]{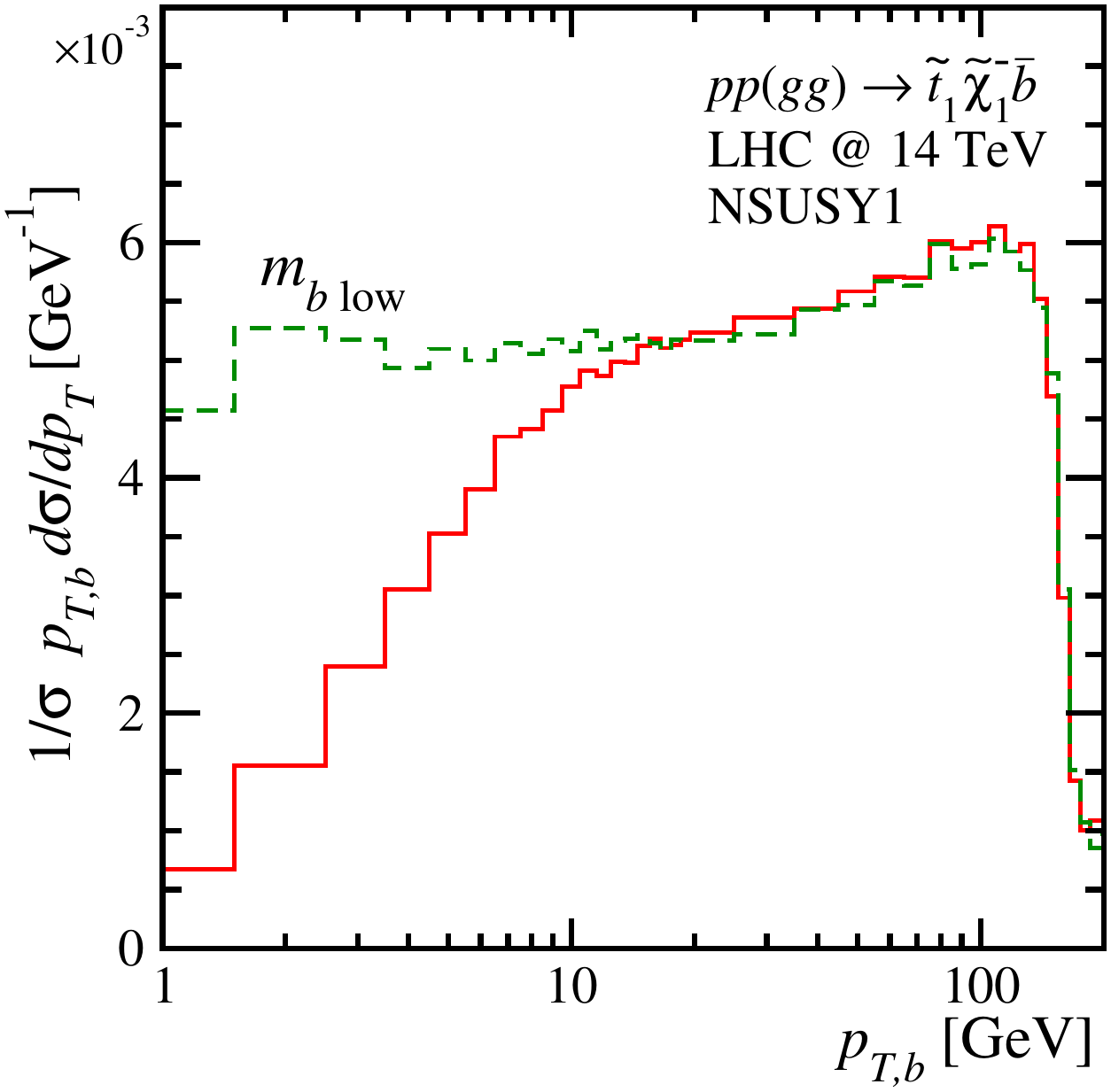} \qquad
 \includegraphics[width=0.3\textwidth]{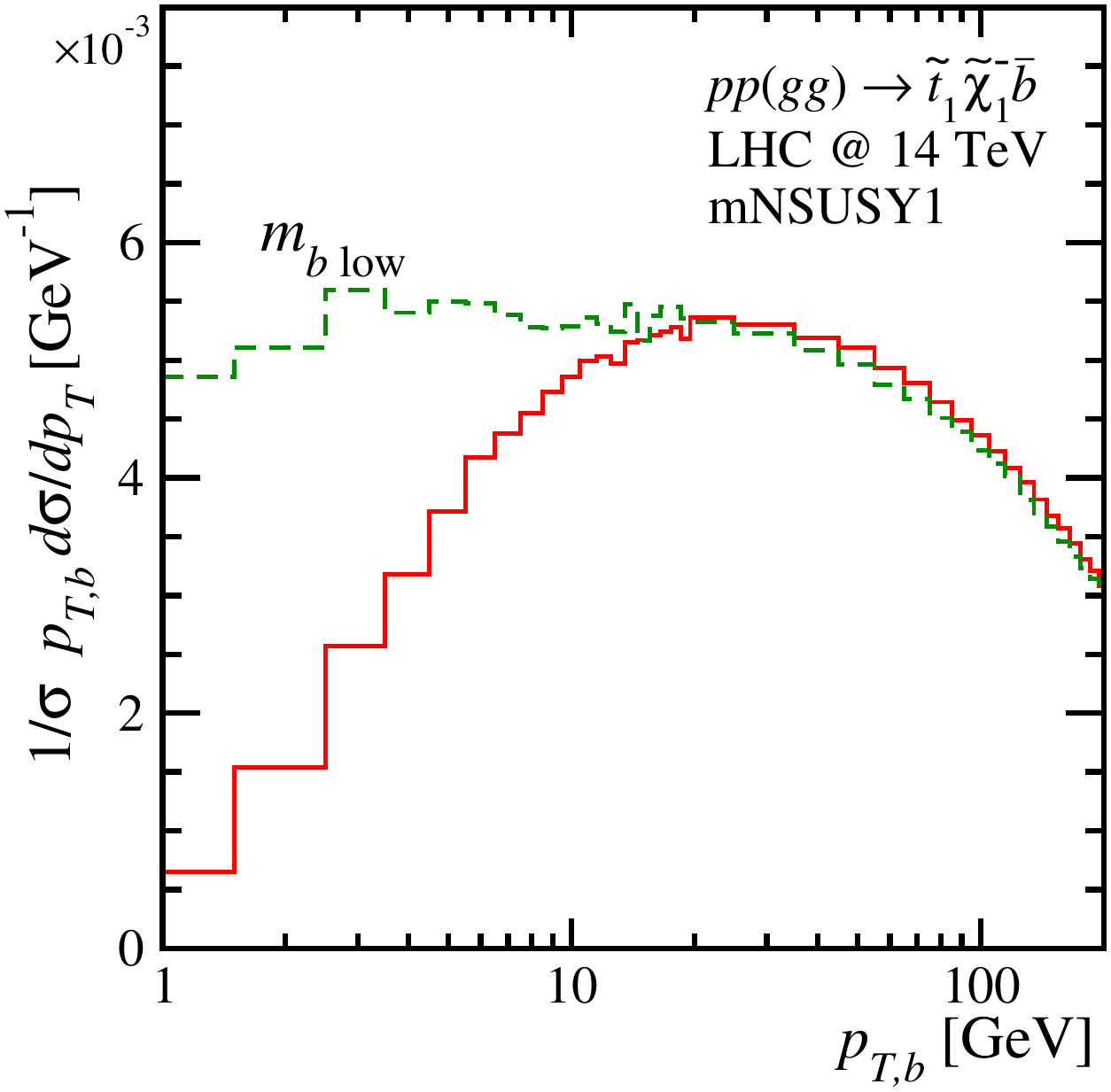} 
\end{center}
\caption{Distributions $p_{T,b}\,d\sigma/dp_{T,b}$ for the process
  $pp(gg)\to\stone\tilde\chi^-_1\bar{b}$ at the 14~TeV LHC.  We
  examine different MSSM benchmarks separately, as described in the
  text.  The (m)NSUSY1 scenario with a low bottom mass, $m_{b\,{\rm
      low}}=m_b/10$, is also shown by a dotted line as a reference.}
\label{fig:4flavtwo}
\end{figure}
%------------------------------------------

In the following we validate the 5-flavor approach, as applied in the
previous Section~\ref{subsec:stopcha}, by analyzing the bottom jet
kinematics  in the 4FS~\cite{4vs5higgs}. As
representative benchmark points, we choose the NSUSY1 scenario defined
in Table~\ref{tab:spectra} and the modified NSUSY1 (mNSUSY1) setup,
which differs from the original NSUSY1 parameter point in the heavier
chargino mass $m_{\tilde\chi^{\pm}_1}=1$~TeV. The latter avoids
resonant stop decays and the associated on--shell resonances.  Here we
use the NNPDF21\_FFN\_NF4\_100~\cite{Ball:2011mu} parton density with
four active flavors and a bottom mass fixed to $m_b=4.78$~GeV.

We show the transverse momentum distribution of the radiated bottom
quark in Figure~\ref{fig:4flavtwo}, normalizing it such that a flat
behavior corresponds to the proper asymptotic scaling shown in
Eq.\eqref{eq:asymptotic}. For the regular parameter point the
distribution has dropped to half of its plateau value for $p_{T,b} \sim
200$~GeV, while for the heavier chargino the distribution extends to
significantly larger transverse momenta. A factorization scale
$\mu_{F,b} = (m_1 + m_2)/2$ appears fully justified by these results,
if we extend the uncertainty band to sufficiently low factorization
scales.

%%%%%%%%%%%%%%%%%%%%%%%%%%%%%%%%%%%%%%%%%%%%%%%%%%%%%%%%%%%%%%%%%%%%%%%%
\section{Summary}
\label{sec:summary}

In the upcoming LHC runs dedicated searches for light stops,
possibly in association with light charginos, will play an
increasingly important role. This requires a precise understanding of
the production processes in perturbative QCD.  We have studied the
pair production and the associated production of heavy--flavor squarks
in the automated \textsc{MadGolem} framework.  For a set of
representative up-to-date MSSM parameter points with light stops and
sbottoms we have examined the total and differential production rates
to NLO accuracy.  We find that

\begin{itemize}
 \item for stop and sbottom pair production the $K$ factors range
   around $1.5 - 1.9$, with significant variations over phase
   space. For the associated stop--chargino case, we find comparably
   constant corrections around $K = 1.1 - 1.5$. In both cases there
   exists a $K$ factor variation over phase space, which might require
   NLO distributions for some phase space regions.
\item these corrections are mainly due to QCD radiation, whilst
  SUSY-QCD corrections are subleading. The NLO distributions agree
  well with predictions from up to two additional jets merged from the
  matrix element and the parton shower.
\item the bottom--inclusive or 5-flavor picture is suitable for
  stop--chargino production, with a bottom factorization scale around
  the average mass of the two heavy states.
\item unlike for stop pairs, the scale dependence of the associated
  production channel cannot be used as a conservative measure of the
  theory uncertainties.
\end{itemize}

Aside from the already built-in \textsc{MadGolem} functionalities,
such as the automated UV renormalization, SUSY dipoles and on--shell
subtraction, the current version has been upgraded to support squark
mixing, the associated counter terms, and finite quark mass effects.
This completes the \textsc{MadGolem} program.\bigskip

\begin{center}
\textbf{Acknowledgments}
\end{center}

It is a pleasure to thank Juan Rojo and Maria Ubiali for kindly
providing us a leading-order version of the NNPDF21\_FFN\_NF4 parton densities
as well as for enlightening discussions.  This work is supported in
part by the Belgian Federal Science Policy Office through the
Interuniversity Attraction Pole P7/37. DLV is funded by the F.R.S.-FNRS \textsl{Fonds de
la Recherche Scientifique} (Belgium).  KM is supported by the
Strategic Research Program \textsl{High Energy Physics} and the Research
Council of the Vrije Universiteit Brussel.\bigskip \bigskip \bigskip

%%%%%%%%%%%%%%%%%%%%%%%%%%%%%%%%%%%%%%%%%%%%%%%%%%%%%%%%%%%%%%%%%%%%%%%%
\appendix
\section{Renormalization}
\label{sec:renormalization}

In this appendix we summarize the aspects of the general MSSM
renormalization relevant for SUSY-QCD interactions involving third
generation squarks.  For a comprehensive account on the notation setup
and the technical implementation in {\sc MadGolem} we refer the reader
to the dedicated appendices of Ref.~\cite{GoncalvesNetto:2012yt}.

The renormalization of the MSSM heavy flavor squarks has been
addressed extensively. The problem was first tackled in the context of
squark decays and was extended later on for squark pair
production~\cite{early-squark}; and radiative corrections to the MSSM
Higgs boson masses~\cite{renorm-higgs}.  Renormalization scheme issues
have been examined in detail
in Refs.~\cite{ren-real,Heinemeyer:2010mm,ren-complex}.

\medskip
\begin{center}\textbf{Leading order parameterization}\end{center}
\smallskip

Let us consider the squark kinetic term within the general MSSM
Lagrangian,
\begin{alignat}{5}
 \lag &= \sum_{\sq = \sqt, \sqb}\,(\partial_\mu \sql^* \, \partial_\mu\sqr^*)\,\left(
\begin{array}{c}
 \partial^\mu\,\sq_L \\ \partial^\mu\,\sq_R 
\end{array}
 \right) - (\sq_L^* \, \sq_R^*)\,\mathcal{M}_{\sq}^2\, 
\left(
\begin{array}{c}
\sq_L \\ \sq_R 
\end{array}
 \right)
\label{eq:kin},
\end{alignat}
with the mass--squared matrix,
\begin{equation}
 \mathcal{M}^2_{\sq} = 
\left( \begin{array}{cc}
       M^2_{\sq_L} + m_q^2 + m_Z^2\,(T_{3, q} - Q_q\swd)\cos 2\beta & m_q\,X_q \\
       m_q\,X_q & M^2_{\sq_R} + m_q^2 + m_Z^2 Q_q  \swd  \cos 2\beta                
      \end{array}
\right)
\label{eq:squarkmass}.
\end{equation}
Following standard conventions, we introduce
$X_q \equiv A_q - \mu\{\cot\beta, \tan\beta\}$ with $\{\cot\beta, \tan\beta\}$
related to $\{t,b\}$ respectively. This parameter is connected to the vacuum expectation
values of the two MSSM Higgs doublet fields, $\tan\beta \equiv v_2/v_1$.
The squark mass terms $M^2_{\sq_{L(R)}}$ and 
the trilinear coupling $A_q$ define the relevant soft-SUSY breaking parameters;
$\mu$ represents the supersymmetric Higgsino mass term; and $m_q, Q_q, T_{3,q}$ stand 
for the (heavy) quark mass, electric charge and third component weak isospin. 

The symmetric mass matrix in Eq.\eqref{eq:squarkmass} is
diagonalized through the unitary transformation $\rotsq \equiv R(\theta_{\sq})$
which rotates the squark gauge eigenstates $\sq_{L(R)}$ into the physical 
squark basis $\sq_{1(2)}$. The squark mass eigenvalues $m_{\sq_{1(2)}}$ and
the mixing angle $\theta_{\sq}$ can thus be expressed in
terms of the MSSM input
parameters, cf. e.g. \cite{sfermionsector}. 

\bigskip
\begin{center}\textbf{Renormalization scheme}\end{center}
\smallskip

The required vertex counterterms are generated by multiplicative
renormalization of the respective wave functions
and model parameters,
\begin{alignat}{5}
\Psi^{(0)} \to Z^{1/2}_{\Psi}\,\Psi \;,
\qqqquad 
m_\Psi^{(0)}  \to m_\Psi + \delta\,m_\Psi \;,
\qqqquad 
g_s^{(0)} \to g_s + \delta\,g_s \; .
\label{eq:defRC}
\end{alignat}
In line with this general procedure, we replace the bare squark field and squark
mass matrices by their respective 
renormalized expressions,
\begin{alignat}{5}
\left( \begin{array}{c}
\sq_1 \\ \sq_2
 \end{array} \right) \; &\rightarrow \; (R^{\sq})\,\left[\mathbb{1}_{2\times 2} + \cfrac{1}{2}\,\delta\,\mathcal{Z}^{\sq}\right]\,
\left( \begin{array}{c}
\sq_L \\ \sq_R
 \end{array} \right)\; \qquad \text{with} \; \qquad \delta\,\mathcal{Z}^{\sq} = 
\left( \begin{array}{cc}
\delta Z_{\sq_1} & \delta Z_{12}^{\sq} \\
\delta Z_{21}^{\sq} & \delta Z_{\sq_2}
\end{array}\right) \;, \notag \\
 R^{\sq}\,\mathcal{M}^2_{\sq}\,(R^{\sq})^\dagger 
 &\rightarrow 
R^{\sq}\,\mathcal{M}^2_{\sq}\,(R^{\sq})^\dagger + R^{\sq}\,\delta\mathcal{M}^2_{\sq}\,(R^{\sq})^\dagger
= \left( 
\begin{array}{cc}
 m^2_{\sq_1} & 0 \\ 0 & m^2_{\sq_2}
\end{array}
\right) + \left( \begin{array}{cc} \delta m^2_{\sq_1} & \delta\,Y^2_{\sq} \\ 
\delta\,Y^2_{\sq} & \delta\,m^2_{\sq_2} \end{array}
\right), 
\label{eq:squarkmassren}
\end{alignat}
where $\delta\,Y^2_{\sq} \equiv 
(m^2_{\sq_1}-m^2_{\sq_2})\,\delta\theta_{\tilde{q}}$. 
Expanding the squark kinetic term of Eq.\eqref{eq:kin} gives for each of the squarks
\begin{alignat}{5}
 \delta\lag = \cfrac{1}{2} \; (\partial_\mu \sq_1^*, \partial_\mu\sq^*_2)
(\delta \mathcal{Z}^{\sq \dagger} + \delta \mathcal{Z}^{\sq})
\left(
\begin{array}{c}
 \partial^\mu \sq_1 \\ \partial^\mu \sq_2 
\end{array}
 \right) - \cfrac{1}{2} (\sq_1^*\, \sq_2^*) 
\left[ \left( \delta \mathcal{Z}^{\sq \dagger} + \delta \mathcal{Z}^{\sq} \right) 
\left(\begin{array}{cc} 
m^2_{\sq_1} & 0 \\ 0 & m^2_{\sq_2}
\end{array} \right) 
+ 2 \left(\begin{array}{cc}  \delta m^2_{\sq_1} & \delta Y^2_{\sq}
\\ \delta Y^2_{\sq} & \delta m^2_{\sq_2} \end{array}\right)\right]
\left( \begin{array}{c} \sq_1 \\ \sq_2 \end{array} \right)
\label{eq:kinren},
\end{alignat}
where from we can readily extract the coefficients 
of the (matrix-valued) renormalized
squark self-energy:
\begin{alignat}{5}
 \retildehat_{\sq_a}(p^2) &= \retilde_{\sq_a}(p^2)+\frac{1}{2}\,
\left(\delta\,Z_{\sq_a} + \delta Z_{\sq_a}^\dagger \right)\,(p^2-m^2_{\sq_a})\,
 - \delta m^2_{\sq_a} \qquad [a = 1,2] \;,
\notag \\
\retildehat_{12}^{\sq}(p^2) &= 
\retilde_{12}^{\sq}(p^2) + \frac{1}{2}\,\delta\,Z^{\sq}_{12}\,(p^2-m^2_{\sq_1})
 + \frac{1}{2}\,\delta\,Z^{\sq}_{21}\,(p^2-m^2_{\sq_2})-
(\delta\,Y^2_{\sq}) \; .
\label{eq:squarkself1}
\end{alignat}

As long as we limit ourselves to SUSY-QCD interactions, the top--stop
system is fully characterized at one loop by four independent
parameters, these are: the top quark mass $m_t$; the top squark masses
$m_{\stop_1}, m_{\stop_2}$; and the top squark mixing angle
$\theta_{\stop}$.  We fix the corresponding counterterms using
generalized on--shell
conditions~\cite{sfermionsector,Heinemeyer:2010mm,thesis} which, 
together with the
mixed wave--function renormalization,
absorb stop mixing in external legs\footnote{Another possibility
is to define a running mixing angle and fully diagonal wave functions ~\cite{thesis}.}:
\begin{enumerate}
 \item on-shell top quark mass
\begin{alignat}{5}
\retildehat_t(m^2_t) = 0 
\quad \rightarrow \quad 
\frac{\delta m_t}{m_t} &= \frac{1}{2}\,\left[\retilde_{t_L}(m_t^2) + \retilde_{t_R}(m_t^2) + 2\retilde_{t_S}(m_t^2)\right]
\label{eq:ren-topmass},
\end{alignat}
where the conventional Lorenz decomposition of the fermionic self-energies reads
\begin{alignat}{5}
 \retilde_q(p^2) = \slashed{p}\,P_L\,\retilde_{q_L}(p^2) + \slashed{p}\,P_R\,\retilde_{q_R}(p^2) + m_f\retilde_{q_S}(p^2) \; .
\label{eq:self-fermion}
\end{alignat}

\item on-shell top squark masses $(a=1,2)$
\begin{alignat}{5}
 \retildehat_{\tilde{t}_a}(m^2_{\tilde{t}_a}) = 0 
\quad \rightarrow \quad \delta m^2_{\tilde{t}_a} = \retilde_{\tilde{t}_a}(m^2_{\stop_a}) \; .
%\\
%%
\label{eq:ren-stopmass}
\end{alignat}

\item no--mixing condition for on-shell stops
\begin{equation}
\retildehat^{\stop}_{12}(m^2_{\stop_a}) = 0 
\label{eq:ren-mixing} \; . 
\end{equation}
In view of Eq.\eqref{eq:squarkself1} 
the latter condition fixes the mixing angle counterterm $\theta_{\stop}$ as:
\begin{alignat}{5}
 \delta\,\theta_{\stop} &= \frac{1}{2}\,\cfrac{\retilde^{\stop}_{12}(m^2_{\stop_1})+\retilde^{\stop}_{12}(m^2_{\stop_2})}{m^2_{\stop_1}-m^2_{\stop_2}}
\quad \text{or} \quad \delta\,Y^2_{\stop} = \frac{1}{2}\,\left[\retilde^{\stop}_{12}(m^2_{\stop_1})+\retilde^{\stop}_{12}(m^2_{\stop_2})\right]
\label{eq:ren-mixing2}.
\end{alignat}
The non--diagonal field renormalization constants then read
\begin{alignat}{5}
\delta Z^{\stop}_{12} = 
\delta Z^{\stop}_{21} = \cfrac{\retilde^{\stop}_{12}(m^2_{\stop_2})-\retilde^{\stop}_{12}(m^2_{\stop_1})}{m^2_{\stop_1}-m^2_{\stop_2}} \; .
\end{alignat}

\item Last, additional on-shell conditions fix the (diagonal) stop
field renormalization constants $(a=1,2)$
\begin{alignat}{5}
 \retildehat'_{\stop_a}(p^2)\Bigg{\lvert}_{p^2 = m^2_{\stop_a}} = 0 \qquad \rightarrow \qquad
 \delta Z_{\stop_a} = -\retilde'_{\stop_a}(m^2_{\stop_a}) \; ,
\label{eq:stopfield}
\end{alignat}
with the conventional shorthand notation $\retilde' \equiv d^2/dp^2 \retilde(p^2)$.
\end{enumerate}

Similar on--shell conditions are used to renormalize the bottom--sbottom sector.
The gauge link between the sbottom and stop sectors 
must be accounted for properly \cite{Heinemeyer:2010mm}.
% 
% %
In practice,  all input masses
in {\sc MadGolem} are defined on--shell. The physical on--shell squark masses
for example from \textsc{SoftSUSY} 
can then be used as input parameters consistent with the on--shell 
squark mass counterterms.

\medskip
\begin{center}\textbf{Renormalization constants}\end{center}
\smallskip

The different field and mass renormalization constants are derived from 
the one--loop self--energies involving the interchange of virtual gluons
and gluinos. 
The strong coupling constant is renormalized in the $\msbar$
scheme explicitly decoupling all particles heavier than the bottom
quark. This zero-momentum subtraction
scheme~\cite{squarkgluinoNLO,Berge:2007dz,decoup} leaves us with the
renormalization group running of $\alpha_s$ from light colored particles
only. It corresponds to the measured value of the strong coupling, for
example in a combined fit with the parton densities. Its
renormalization constant reads
\begin{alignat}{5}
\delta\,g_s = 
- \frac{\alpha_s}{4\pi} \;
  \frac{\beta^L_0+\beta^H_0}{2} \, \frac{1}{\tilde{\epsilon}}
- \frac{\alpha_s}{4\pi} \,
\left( \frac{1}{3} \, \log \frac{m^2_t}{\mu_R^2}
      + \log \frac{m^2_{\go}}{\mu_R^2}
      + \frac{1}{12} \, \sum_\text{squarks} \,
        \log \frac{m^2_{\sq_a}}{\mu_R^2}
\right) \; .
\label{eq:alphas_ct}
\end{alignat}
The UV divergence appears as $1/\tilde{\epsilon} \equiv
(4\pi)^{\epsilon}/\Gamma(1-\epsilon) = 1/\epsilon - \gamma_E +
\log(4\pi) + \mathcal{O}(\epsilon)$.  Both light ($L$) and heavy
($H$) colored particles contribute to the beta function
\begin{alignat}{5}
\beta_0 = 
\beta_0^L + \beta_0^H = 
\left[\frac{11}{3}\,C_A - \frac{2}{3}\,n_f \right]
+ 
\left[- \frac{2}{3}  -\frac{2}{3}\,C_A - \frac{1}{3}\,(n_f+1) \right] \; . 
\end{alignat}
\mg sets the number of active flavors to $n_f = 5$.  The $SU(3)_C$
color factors are $C_F = 4/3$ and $C_A = 3$.\bigskip

In the case
of the gluon field, the heavy colored particles are decoupled consistently
with the 5-flavor scheme used for the strong coupling constant renormalization.
The underlying Slavnov-Taylor identities relate the corresponding finite
counterterms as $\delta Z_G = -2\delta g_s$.
Explicit analytical expressions for the field and mass
renormalization constants relevant for the processes considered
in this paper are documented below. The 
results are given in terms of standard 
Passarino-Veltman scalar functions~\cite{oneloop}

\begin{itemize}
\item[--] (massive) quark field
\begin{alignat}{5}
 \delta Z_{q_{L/R}} &=
\frac{\alpha_s\,C_F}{4\pi}\,
\Big{\{}1 + 4m_q^2\,B_0'(m_q^2,0,m_q^2) + 2\,m_{\go}\,m_q\,\sin(2\theta_{\tilde{q}})\,
\left[B_0'(m_q^2,m_{\sq_1}^2,m_{\go}^2) -  B_0'(m_q^2,m_{\sq_2}^2,m_{\go}^2) \right]  \notag \\ 
& +\cfrac{\cos^2\theta_{\tilde{q}}\,A_0(m^2_{\sq_1}) + \sin^2\theta_{\tilde{q}}\,A_0(m^2_{\sq_2}) -A_0(m_q^2)-A_0(m^2_{\go})}{m_q^2}
- \frac{m^2_{\sq_1} - m^2_{\go}}{m^2_{q}}\,B_0(m_q^2,m^2_{\go},m^2_{\sq_1})  \notag \\
& \left. - \frac{m^2_{\sq_2} - m^2_{\go}}{m^2_{q}}\,B_0(m_q^2,m^2_{\go},m^2_{\sq_2})
-(m_q^2+m^2_{\go}-m^2_{\sq_1})\,\left[ B_0'(m_q^2,m^2_{\go},m^2_{\sq_1}) + \frac{\sin^2\theta_{\sq}}{m_q^2}
\,B_0(m_q^2,m^2_{\go},m^2_{\sq_1}) \right] \right. \notag \\
&  -(m_q^2+m^2_{\go}-m^2_{\sq_2})\,\left[ B_0'(m_q^2,m^2_{\go},m^2_{\sq_2}) + \frac{\cos^2\theta_{\sq}}{m_q^2}
\,B_0(m_q^2,m^2_{\go},m^2_{\sq_2})
 \right]\Big{\}} \; .
\label{eq:quarkfield-rc}
\end{alignat}

\item[--] (massive) quark mass
\begin{alignat}{5}
 \frac{\delta\, m_q}{m_q} &= - \frac{\alpha_s\,C_F}{8\pi\,m_q^2}\,
\Big{\{} 2m_q^2 + 6A_0(m_q^2)+2 A_0(m_{\go}^2)-A_0(m^2_{\sq_1}) - A_0(m^2_{\sq_2}) \notag \\ 
& -2m_q m_{\go}\,\sin (2\theta_{\sq})\,\left[
B_0(m_q^2,m_{\go}^2,m_{\sq_2}^2) - B_0(m_q^2,m_{\go}^2,m_{\sq_1}^2) \right]\,
\notag \\
&    -(m_q^2-m^2_{\sq_1}+m_{\go}^2)\,B_0(m_q^2,m_{\go}^2,m^2_{\sq_1}) 
-(m_q^2-m^2_{\sq_2}+m_{\go}^2)\,B_0(m_q^2,m_{\go}^2,m^2_{\sq_2})\Big{\}} \; .
\label{eq:massq-dimreg} 
\end{alignat}
% %
% 
\item[--] squark field (diagonal)
\begin{alignat}{5}
\label{eq:squarkfield-rc} 
 \delta\,Z_{\sq_1} &= -\frac{\alpha_s\,C_F}{2\pi}\,
\left[ 
B_0(m^2_{\sq},m_q^2,m^2_{\go}) - B_0(m^2_{\sq_1},0,m^2_{\sq_1})\, \right. \\
&+ \left.2m_{\go}\,m_q\,\sin (2\theta_{\sq})\,B_0^{'}(m_{\sq_1}^2,m_q^2,m^2_{\go})
-(m_q^2+m_{\go}^2-m_{\tilde{q}}^2)\,B_0^{'}(m_{\sq_1}^2,m_q^2,m^2_{\go}) - 2m_{\sq_1}^2\,B_0^{'}(m^2_{\sq_1},0,m^2_{\sq_1})
\right] \notag \, ,
\end{alignat}
and likewise for $\sq_2$ with $1\to 2$ and $\theta_{\sq} \to -\theta_{\sq}$.

\item[--] squark field (mixing)
\begin{alignat}{5}
 \delta\,Z_{12}^{\sq} = -\frac{\alpha_s\,C_F}{\pi}\,\frac{m_q\,m_{\go}}{(m^2_{\sq_1}-m^2_{\sq_2})}\,
\cos(2\theta_{\sq})\,\left[B_0(m^2_{\sq_1},m^2_q,m^2_{\go}) -
 B_0(m^2_{\sq_2},m^2_q,m^2_{\go}) \right] \;.
\label{eq:dz12} 
\end{alignat}

\item[--] squark mixing angle
\begin{alignat}{5}
 \delta\,\theta_{\sq} &= \frac{\alpha_S\,C_F}{4\pi}\,\frac{\cos\,(2\theta_{\sq})}{\mssqo-\mssqt}\,
\Big{\{}2 m_b m_{\go}\, \left[B_0(\mssqo, m^2_q, m^2_{\go}) + B_0(\mssqt, m^2_q, m^2_{\go})\right]
+ \sin (2\theta_{\sq})\,\left[A_0(\mssqt) - A_0(\mssqo) \right] \Big{\}} \;.
\label{dzangle}
\end{alignat}

\item[--] squark mass
\begin{alignat}{5}
\delta m^2_{\sq_1} &=  
-\frac{\alpha_s\,C_F}{4\pi}\, \left[ 4m^2_{\sq_1} + 
2\,A_0(m^2_{q}) + 2\,A_0(m^2_{\go}) + 2\,A_0(m^2_{\sq_1}) \right.  + 2(m^2_{\go} - m^2_{\sq_1} + m^2_{q})\,
B_0(m^2_{\sq_1}, m^2_{\go}, m^2_{q})   \notag \\
&  \left. - 4m_{q}\,\,m_{\go}\,\sin(2\theta_{\tilde{q}})\,B_0(m^2_{\sq_1},m^2_{\go}, m^2_{q}) 
+ \sin^2 (2\theta_{\tilde{q}}) \, \left( A_0(m^2_{\sq_1}) -  A_0(m^2_{\sq_2})\right)
\right] \; ,
\label{eq:massrc}
\end{alignat}
and likewise for $\sq_2$ with $1\to 2$ and $\theta_{\sq} \to -\theta_{\sq}$. 

\end{itemize}

\medskip
\begin{center}\textbf{Counterterms}\end{center}
\smallskip

The strong interaction counter terms for squark pair production in the MSSM 
are given by:
\begin{tabular}{cl} & \\ & \\
\myrbox{\includegraphics[width=0.15\textwidth]{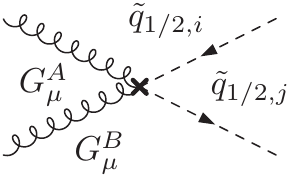}} & 
  $-i\,g_s\,T^A_{ij}\,\left[
    \delta\,g_s + \dfrac{\delta\,Z_{\sq_{1/2,i}}+\delta\,Z_{\sq_{1/2, j}} + \delta\,Z_G}{2} \right]\,
    \sq_{1/2, i}\,(p_i + p_j)^\mu\,G^A_\mu\,\sq_{1/2, j} $  
\\ & \\ 
\myrbox{\includegraphics[width=0.15\textwidth]{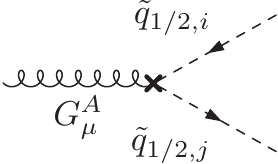}} &
  $i\,g^2_s\,\{T^A\,T^B\}_{ij}\,
  \left[2\delta\,g_s + \delta\,Z_{G} + \dfrac{\delta\,Z_{\sq_{1/2,i}}+\delta\,Z_{\sq_{1/2, j}}}{2} \right] 
  \, \sq_{1/2, i}\,\sq_{1/2, j}\,G_\mu^A\,G^{B\mu}$
  \\ & \\
\end{tabular}
\bigskip

The SUSY-electroweak interactions to leading order read
\begin{alignat}{5}
 g_{\stop_a b\chai} &= \tilde{t}_a\,\bar{b}\,\left(\,\gtchalo\,P_R
+ \gtchart\,P_L \right)\,\tilde{\chi}_i^- \;,\qquad  
%%%%%%%%%
 g_{\stop_a b\chaip} &= \tilde{t}^*_a\,\tilde{\chi}_i^+\,\left(\,\gptchalo\,P_R
+ \gtchart\,P_L \right)\,b \; ,
\label{eq:stopcha-lo-one}
\end{alignat}
with coupling components 
\begin{alignat}{5}
g^{(a)}_{\stop_1 b\chai} &= \rtoo \gtchala + \rtot \gtchara \;\qquad
g^{(a)}_{\stop_2 b\chai} = \rtto \gtchala + \rttt \gtchara \;,\notag \\
%%%%%%
g^{(a)}_{\stop^*_1 b\chai} &= \rtoo \gptchala + \rtot \gptchara \;,\qquad
g^{(a)}_{\stop^*_2 b\chai} = \rtto \gptchala + \rttt \gptchara  \quad 
(a= 1,2) \;,  
\label{eq:stopcha-lo-two}
\end{alignat}
and
\begin{alignat}{5}
 \gtchalo = \gptchalt &= \cfrac{-eV_{11}}{\sw}\;, \qquad 
 \gtchalt = \gptchalo = \cfrac{em_bU_{12}}{\sqrt{2}\,\sw\,m_W\,\cos\beta} \;,\notag \\
\gtcharo = \gptchart &= \cfrac{em_tV_{12}}{\sqrt{2}\,\sw\,m_W\,\sin\beta}\;, \qquad \gtchart = \gptcharo = 0 \; .
\label{eq:stopcha-lo-three}
\end{alignat}
in terms of the chargino mixing matrices $U$ and $V$.  The finite
SUSY--restoring counterterm $\delta_{\text{SUSY}} = -\alpha_s/(6\pi)$
corrects the mismatch of two gaugino and the $2-2\epsilon$ degrees of
freedom induced by the use of dimensional regularization. The
corresponding counter terms are

\begin{tabular}{cl} 
 & \\ & \\
\multirow{5}{3cm}{\myrbox{\includegraphics[width=0.15\textwidth]{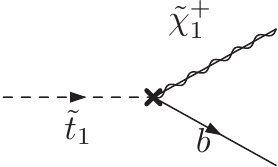}}} &
$\delta[g^{(1)}_{\stop_1b\chai}] = \gtchalo\,\left[\rtoo\,\left(\cfrac{\delta Z_{b_L}^\dagger+\delta Z_{\stop_1}}{2} + \delta_{\text{SUSY}} \right) 
+ \rtto\, \cfrac{\delta Z^{\stop}_{12}}{2} \right] $ \\ & $\phantom{\delta[g^{(1)}_{\stop_1b\chai}] = \gtchalo\,} + \gtcharo
\left[\rtot\,\left(\cfrac{\delta Z_{b_L}^\dagger+\delta Z_{\stop_1}}{2} + \cfrac{\delta m_t}{m_t} + \delta_{\text{SUSY}}\right) 
+ \rttt\, \cfrac{\delta Z^{\stop}_{12}}{2} \right]$  \\    
\\ & \\ 
%%%%%%%%%%%%%%%%%
 & $\delta[g^{(2)}_{\stop_1b\chai}] = \gtchalt\,\left[\rtoo\,\left(\cfrac{\delta Z_{b_R}^\dagger+\delta Z_{\stop_1}}{2} + \cfrac{\delta m_b}{m_b}+ \delta_{\text{SUSY}} \right) 
+ \rtto\, \cfrac{\delta Z^{\stop}_{12}}{2} \right] $ \\ & $\phantom{\delta[g^{(1)}_{\stop_1b\chai}] = \gtchalo\,} + \gtchart
\left[\rtot\,\left(\cfrac{\delta Z_{b_R}^\dagger+\delta Z_{\stop_1}}{2} + \delta_{\text{SUSY}}\right) 
+ \rttt\, \cfrac{\delta Z^{\stop}_{12}}{2} \right]$  \\    
\\ & \\ 
%%%%%%%%%%%%%%%%
\end{tabular}

\begin{tabular}{cl} 
 & \\ & \\
\multirow{2}{3cm}{\myrbox{\includegraphics[width=0.15\textwidth]{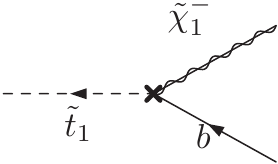}}} &
$\delta[g^{(1)}_{\stop_1b\chai}] = \gtchalo\,\left[\rtoo\,\left(\cfrac{\delta Z_{b_R}+\delta Z_{\stop_1}}{2}  + \cfrac{\delta m_b}{m_b} + \delta_{\text{SUSY}}\right) 
+ \rtto\, \cfrac{\delta Z^{\stop}_{12}}{2} \right] $ \\ & $\phantom{\delta[g^{(1)}_{\stop_1b\chai}] = \gtchalo\,} + \gtcharo
\left[\rtot\,\left(\cfrac{\delta Z_{b_R}+\delta Z_{\stop_1}}{2}\right) 
+ \rttt\, \cfrac{\delta Z^{\stop}_{12}}{2} \right]$  \\
  \\ & \\
& $\delta[g^{(2)}_{\stop_1b\chai}] = \gtchalt\,\left[\rtoo\,\left(\cfrac{\delta Z_{b_L}+\delta Z_{\stop_1}}{2} + \delta_{\text{SUSY}} \right) 
+ \rtto\, \cfrac{\delta Z^{\stop}_{12}}{2} \right] $ \\ & $\phantom{\delta[g^{(1)}_{\stop_1b\chai}] = \gtchalo\,} + \gtchart
\left[\rtot\,\left(\cfrac{\delta Z_{b_L}+\delta Z_{\stop_1}}{2} + \cfrac{\delta m_t}{m_t} + \delta_{\text{SUSY}}\right) 
+ \rttt\, \cfrac{\delta Z^{\stop}_{12}}{2} \right]$  \\
  \\ & \\
%%%%%%%%%%%%%%%%
\multirow{5}{3cm}{\myrbox{\includegraphics[width=0.15\textwidth]{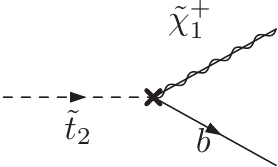}}} &
$\delta[g^{(1)}_{\stop_2b\chai}] = \gtchalo\,\left[\rtto\,\left(\cfrac{\delta Z_{b_L}^\dagger+\delta Z_{\stop_2}}{2}+ \delta_{\text{SUSY}} \right) 
+ \rtoo\, \cfrac{\delta Z^{\stop}_{12}}{2} \right] $ \\ & $\phantom{\delta[g^{(1)}_{\stop_1b\chai}] = \gtchalo\,} + \gtcharo
\left[\rttt\,\left(\cfrac{\delta Z_{b_L}^\dagger+\delta Z_{\stop_2}}{2} + \cfrac{\delta m_t}{m_t} + \delta_{\text{SUSY}}\right) 
+ \rtot\, \cfrac{\delta Z^{\stop}_{12}}{2} \right]$  \\    
\\ & \\ 
%%%%%%%%%%%%%%%%%
 & $\delta[g^{(2)}_{\stop_2b\chai}] = \gtchalt\,\left[\rtto\,\left(\cfrac{\delta Z_{b_R}^\dagger+\delta Z_{\stop_1}}{2} + \cfrac{\delta m_b}{m_b} + \delta_{\text{SUSY}}\right) 
+ \rtoo\, \cfrac{\delta Z^{\stop}_{12}}{2} \right] $ \\ & $\phantom{\delta[g^{(1)}_{\stop_1b\chai}] = \gtchalo\,} + \gtchart
\left[\rttt\,\left(\cfrac{\delta Z_{b_R}^\dagger+\delta Z_{\stop_2}}{2} + \delta_{\text{SUSY}}\right) 
+ \rtot\, \cfrac{\delta Z^{\stop}_{12}}{2} \right]$  \\    
\\ & \\ 
%%%%%%%%%%%%%%%%
%%%%%%%%%%%%%%%%
\end{tabular}

\begin{tabular}{cl} 
\multirow{2}{3cm}{\myrbox{\includegraphics[width=0.15\textwidth]{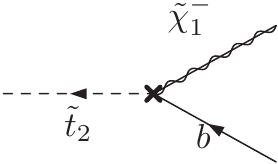}}} &
$\delta[g^{(1)}_{\stop_2b\chai}] = \gtchalo\,\left[\rtto\,\left(\cfrac{\delta Z_{b_R}+\delta Z_{\stop_2}}{2}  + \cfrac{\delta m_b}{m_b} + \delta_{\text{SUSY}}\right) 
+ \rtoo\, \cfrac{\delta Z^{\stop}_{12}}{2} \right] $ \\ & $\phantom{\delta[g^{(1)}_{\stop_1b\chai}] = \gtchalo\,} + \gtcharo
\left[\rttt\,\left(\cfrac{\delta Z_{b_R}+\delta Z_{\stop_2}}{2} + \delta_{\text{SUSY}}\right) 
+ \rtot\, \cfrac{\delta Z^{\stop}_{12}}{2} \right]$  \\
  \\ & \\
& $\delta[g^{(2)}_{\stop_2b\chai}] = \gtchalt\,\left[\rtto\,\left(\cfrac{\delta Z_{b_L}+\delta Z_{\stop_2}}{2} + \delta_{\text{SUSY}} \right) 
+ \rtoo\, \cfrac{\delta Z^{\stop}_{12}}{2} \right] $ \\ & $\phantom{\delta[g^{(1)}_{\stop_1b\chai}] = \gtchalo\,} + \gtchart
\left[\rttt\,\left(\cfrac{\delta Z_{b_L}+\delta Z_{\stop_2}}{2} + \cfrac{\delta m_t}{m_t} + \delta_{\text{SUSY}}\right) 
+ \rtot\, \cfrac{\delta Z^{\stop}_{12}}{2} \right]$  \\
  \\ & \\
\end{tabular}
%------------------------------------------------

Finally, we document the required counterterms
to renormalize the UV divergent 2-point functions
associated to heavy flavor squarks. 
Non--diagonal loop--induced transitions $\tilde{q}_1 - \tilde{q}_2$
give rise to additional divergent structures, as compared
to light generation squarks:

%------------------------------------------------
\begin{tabular}{cl} & \\ & \\
\myrbox{\includegraphics[width=0.15\textwidth]{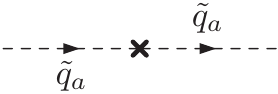}} & 
  $ i\left[ (p^2 - m^2_{\tilde{q}})\,\delta Z_{\tilde{q_a}} - \delta m^2_{\tilde{q}_a}\right]$  
\\ & \\ 
\myrbox{\includegraphics[width=0.15\textwidth]{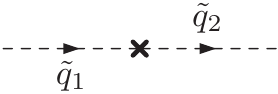}} &
  $ i\left[ p^2 \delta Z^{\tilde{q}}_{12} - \cfrac{1}{2}(m^2_{\tilde{q}_1} + m^2_{\tilde{q}_2})\,\delta Z^{\tilde{q}}_{12} - \delta\,Y^2_{\tilde{q}}\right]$  
 \\ & \\
\end{tabular}
%------------------------------------------------

%%%%%%%%%%%%%%%%%%%%%%%%%%%%%%%%%%%%%%%%%%%%%%%%%%%%%%%%%%%%%%%%%%%%%%%


\begin{thebibliography}{99}

\bibitem{Morrissey:2009tf} 
  D.~E.~Morrissey, T.~Plehn and T.~M.~P.~Tait,
  %``Physics searches at the LHC,''
  Phys.\ Rept.\  {\bf 515}, 1 (2012).
  %[arXiv:0912.3259 [hep-ph]].
  %%CITATION = ARXIV:0912.3259;%%

\bibitem{susy_fits}
 C.~Boehm, P.~S.~B.~Dev, A.~Mazumdar and E.~Pukartas,
  %``Naturalness of Light Neutralino Dark Matter in pMSSM after LHC, XENON100 and Planck Data,''
  JHEP {\bf 1306}, 113 (2013);
  %[arXiv:1303.5386 [hep-ph]].
  %%CITATION = ARXIV:1303.5386;%%
   T.~Cohen and J.~G.~Wacker,
  %``Here be Dragons: The Unexplored Continents of the CMSSM,''
  JHEP {\bf 1309}, 061 (2013);
  %[arXiv:1305.2914 [hep-ph]].
  %%CITATION = ARXIV:1305.2914;%%
  A.~Fowlie, K.~Kowalska, L.~Roszkowski, E.~M.~Sessolo and Y.~-L.~S.~Tsai,
  %``Dark matter and collider signatures of the MSSM,''
  Phys.\ Rev.\ D {\bf 88}, 055012 (2013);
  %[arXiv:1306.1567 [hep-ph]];
  %%CITATION = ARXIV:1306.1567;%%
  S.~Henrot-Versille \etal,
  %, R.~Lafaye, T.~Plehn, M.~Rauch, D.~Zerwas, S.~Plaszczynski, B.~Rouillé d'Orfeuil and M.~Spinelli,
  %``Constraining Supersymmetry using the relic density and the Higgs boson,''
  Phys.\ Rev.\ D {\bf 89}, 055017 (2014);
  %[arXiv:1309.6958 [hep-ph]].
  %%CITATION = ARXIV:1309.6958;%%
  P.~Bechtle \etal,
  %, K.~Desch, H.~K.~Dreiner, M.~Hamer, M.~Krämer, B.~O'Leary, W.~Porod and X.~Prudent {\it et al.},
  %``Constrained Supersymmetry after the Higgs Boson Discovery: A global analysis with Fittino,''
  arXiv:1310.3045 [hep-ph];
  %%CITATION = ARXIV:1310.3045;%%
  %15 citations counted in INSPIRE as of 30 Jun 2014
  O.~Buchmueller \etal. 
  %, R.~Cavanaugh, A.~De Roeck, M.~J.~Dolan, J.~R.~Ellis, H.~Flacher, S.~Heinemeyer and G.~Isidori {\it et al.},
  %``The CMSSM and NUHM1 after LHC Run 1,''
  arXiv:1312.5250 [hep-ph];
  %%CITATION = ARXIV:1312.5250;%%
  J.~Ellis, K.~A.~Olive and J.~Zheng,
  %``The Extent of the Stop Coannihilation Strip,''
  arXiv:1404.5571 [hep-ph].
  %%CITATION = ARXIV:1404.5571;%%

\bibitem{searches-gen}
 %\cite{Aad:2012fqa}
% \bibitem{Aad:2012fqa} 
  G.~Aad {\it et al.}  [ATLAS Collaboration],
  %``Search for squarks and gluinos with the ATLAS detector in final states with jets and missing transverse momentum using 4.7 fb^{-1}$ of $\sqrt{s}=7$ TeV proton-proton collision data,''
  Phys.\ Rev.\ D {\bf 87}, 012008 (2013)
  [arXiv:1208.0949 [hep-ex]]; %\cite{Chatrchyan:2012lia}
% \bibitem{Chatrchyan:2012lia} 
  S.~Chatrchyan {\it et al.}  [CMS Collaboration],
  %``Search for new physics in the multijet and missing transverse momentum final state in proton-proton collisions at $\sqrt{s} = 7$ TeV,''
  Phys.\ Rev.\ Lett.\  {\bf 109}, 171803 (2012)
  [arXiv:1207.1898 [hep-ex]]; and the most recent (not yet published) updates, cf. e.g. CMS-PAS-SUS-12-028
  %%CITATION = ARXIV:1207.1898;%%

\bibitem{atlassite}
\texttt{https://twiki.cern.ch/twiki/bin/view/AtlasPublic/SupersymmetryPublicResults}

\bibitem{cmssite}
\texttt{https://twiki.cern.ch/twiki/bin/view/CMSPublic/PhysicsResultsSUS}

\bibitem{nsusy} see e.g. 
  B.~de Carlos and J.~A.~Casas,
  %``One loop analysis of the electroweak breaking in supersymmetric models and the fine tuning problem,''
  Phys.\ Lett.\ B {\bf 309}, 320 (1993); 
  %\cite{Dine:1993np}
% \bibitem{Dine:1993np} 
  M.~Dine, R.~G.~Leigh and A.~Kagan,
  %``Flavor symmetries and the problem of squark degeneracy,''
  Phys.\ Rev.\ D {\bf 48}, 4269 (1993); %\cite{Anderson:1994dz}
  G.~W.~Anderson and D.~J.~Castano,
  %``Measures of fine tuning,''
  Phys.\ Lett.\ B {\bf 347}, 300 (1995);
  S.~Dimopoulos and G.~F.~Giudice,
  %``Naturalness constraints in supersymmetric theories with nonuniversal soft terms,''
  Phys.\ Lett.\ B {\bf 357}, 573 (1995);
  A.~G.~Cohen, D.~B.~Kaplan and A.~E.~Nelson,
  %``The More minimal supersymmetric standard model,''
  Phys.\ Lett.\ B {\bf 388}, 588 (1996);
  R.~Kitano and Y.~Nomura,
  %``Supersymmetry, naturalness, and signatures at the LHC,''
  Phys.\ Rev.\ D {\bf 73}, 095004 (2006);
  C.~Brust, A.~Katz, S.~Lawrence and R.~Sundrum,
  %``SUSY, the Third Generation and the LHC,''
  JHEP {\bf 1203}, 103 (2012);
  H.~M.~Lee, V.~Sanz and M.~Trott,
  %``Hitting sbottom in natural SUSY,''
  JHEP {\bf 1205}, 139 (2012);
  %[arXiv:1204.0802 [hep-ph]].
  %%CITATION = ARXIV:1204.0802;%%
  M.~Papucci, J.~T.~Ruderman and A.~Weiler,
  %``Natural SUSY Endures,''
  JHEP {\bf 1209}, 035 (2012);
  J.~R.~Espinosa, C.~Grojean, V.~Sanz and M.~Trott,
  %``NSUSY fits,''
  JHEP {\bf 1212}, 077 (2012);
  K.~Kowalska and E.~M.~Sessolo,
  %``Natural MSSM after the LHC 8 TeV run,''
  arXiv:1307.5790 [hep-ph];
  %%CITATION = ARXIV:1307.5790;%% 
  F.~Br\"ummer, S.~Kraml, S.~Kulkarni and C.~Smith,
  %``The Flavour of Natural SUSY,''
  arXiv:1402.4024 [hep-ph].
  %%CITATION = ARXIV:1402.4024;%%
  %1 citations counted in INSPIRE as of 04 Apr 2014

\bibitem{nsusy-benchmarks}
  H.~Baer, V.~Barger, P.~Huang and X.~Tata,
  %``Natural Supersymmetry: LHC, dark matter and ILC searches,''
  JHEP {\bf 1205}, 109 (2012); 
  %[arXiv:1203.5539 [hep-ph]].
  %%CITATION = ARXIV:1203.5539;%%
  H.~Baer and J.~List,
  %``Post LHC8 SUSY benchmark points for ILC physics,''
  Phys.\ Rev.\ D {\bf 88}, 055004 (2013).
  %[arXiv:1307.0782 [hep-ph]].
  %%CITATION = ARXIV:1307.0782;%%
  %5 citations counted in INSPIRE as of 30 Apr 2014

\bibitem{Buchmueller:2013exa} 
  O.~Buchmueller and J.~Marrouche,
  %``Universal mass limits on gluino and third-generation squarks in the context of Natural-like SUSY spectra,''
  Int.\ J.\ Mod.\ Phys.\ A {\bf 29}, 1450032 (2014).
  %[arXiv:1304.2185 [hep-ph]].
  %%CITATION = ARXIV:1304.2185;%%
  %11 citations counted in INSPIRE as of 30 Apr 2014

\bibitem{Brummer:2012ns} 
  F.~Br\"ummer, S.~Kraml and S.~Kulkarni,
  %``Anatomy of maximal stop mixing in the MSSM,''
  JHEP {\bf 1208}, 089 (2012).
  %[arXiv:1204.5977 [hep-ph]].
  %%CITATION = ARXIV:1204.5977;%%
  %45 citations counted in INSPIRE as of 24 Jul 2013

\bibitem{Han:2013mga} 
  T.~Han, T.~Li, S.~Su and L.~-T.~Wang,
  %``Non-Decoupling MSSM Higgs Sector and Light Superpartners,''
  arXiv:1306.3229 [hep-ph].
  %%CITATION = ARXIV:1306.3229;%%

\bibitem{stop_coann} 
 J.~R.~Ellis, K.~A.~Olive and Y.~Santoso,
  %``Calculations of neutralino stop coannihilation in the CMSSM,''
  Astropart.\ Phys.\  {\bf 18}, 395 (2003);
  %[hep-ph/0112113].
  %%CITATION = HEP-PH/0112113;%%
  C.~Boehm, A.~Djouadi and M.~Drees,
  %``Light scalar top quarks and supersymmetric dark matter,''
  Phys.\ Rev.\ D {\bf 62}, 035012 (2000).
  %[hep-ph/9911496].
  %%CITATION = HEP-PH/9911496;%%
  %170 citations counted in INSPIRE as of 15 Apr 2014

\bibitem{Morrissey:2012db} 
  D.~E.~Morrissey and M.~J.~Ramsey-Musolf,
  %``Electroweak baryogenesis,''
  New J.\ Phys.\  {\bf 14}, 125003 (2012).
  %[arXiv:1206.2942 [hep-ph]].
  %%CITATION = ARXIV:1206.2942;%%
  %27 citations counted in INSPIRE as of 24 Jul 2013

\bibitem{stop-atlas-papers}
%\cite{Aad:2012ywa}
% \bibitem{Aad:2012ywa} 
  G.~Aad {\it et al.}  [ATLAS Collaboration],
  %``Search for a supersymmetric partner to the top quark in final states with jets and missing transverse momentum at $\sqrt{s}=7$ TeV with the ATLAS detector,''
  Phys.\ Rev.\ Lett.\  {\bf 109}, 211802 (2012);
%   [arXiv:1208.1447 [hep-ex]].
  %%CITATION = ARXIV:1208.1447;%%
  %49 citations counted in INSPIRE as of 07 Aug 2013
%\cite{Aad:2012xqa}
% \bibitem{Aad:2012xqa} 
  G.~Aad {\it et al.}  [ATLAS Collaboration],
  %``Search for direct top squark pair production in final states with one isolated lepton, jets, and missing transverse momentum in $\sqrt{s}=7$ TeV $pp$ collisions using 4.7 $fb^{-1}$ of ATLAS data,''
  Phys.\ Rev.\ Lett.\  {\bf 109}, 211803 (2012);
%   [arXiv:1208.2590 [hep-ex]].
  %%CITATION = ARXIV:1208.2590;%%
  %44 citations counted in INSPIRE as of 07 Aug 2013
%\cite{Aad:2012uu}
% \bibitem{Aad:2012uu} 
  G.~Aad {\it et al.}  [ATLAS Collaboration],
  %``Search for a heavy top-quark partner in final states with two leptons with the ATLAS detector at the LHC,''
  JHEP {\bf 1211}, 094 (2012);
%   [arXiv:1209.4186 [hep-ex]].
  %%CITATION = ARXIV:1209.4186;%%
  %30 citations counted in INSPIRE as of 07 Aug 2013
%\cite{Aad:2012tx}
% \bibitem{Aad:2012tx} 
  G.~Aad {\it et al.}  [ATLAS Collaboration],
  %``Search for light scalar top quark pair production in final states with two leptons with the ATLAS detector in $\sqrt{s}=7$ TeV proton-proton collisions,''
  Eur.\ Phys.\ J.\ C {\bf 72}, 2237 (2012);
%   [arXiv:1208.4305 [hep-ex]].
  %%CITATION = ARXIV:1208.4305;%%
  %27 citations counted in INSPIRE as of 07 Aug 2013
%\cite{Aad:2012yr}
% \bibitem{Aad:2012yr} 
  G.~Aad {\it et al.}  [ATLAS Collaboration],
  %``Search for light top squark pair production in final states with leptons and $b^-$ jets with the ATLAS detector in $\sqrt{s}=7$ TeV proton-proton collisions,''
  Phys.\ Lett.\ B {\bf 720}, 13 (2013).
%   [arXiv:1209.2102 [hep-ex]].
  %%CITATION = ARXIV:1209.2102;%%
  %34 citations counted in INSPIRE as of 07 Aug 2013


\bibitem{stop-atlas-conf} 
ATLAS-CONF-2013-024, ATLAS-CONF-2013-037; ATLAS-CONF-2013-053; ATLAS-CONF-2013-065

\bibitem{stop-cms-conf}  CMS-PAS-SUS-12-023; CMS-PAS-SUS-12-028; CMS-PAS-SUS-13-011

\bibitem{light_stop_para}
  X.~-J.~Bi, Q.~-S.~Yan and P.~-F.~Yin,
  %``Probing Light Stop Pairs at the LHC,''
  Phys.\ Rev.\ D {\bf 85}, 035005 (2012);
  %[arXiv:1111.2250 [hep-ph]].
  %%CITATION = ARXIV:1111.2250;%%
  N.~Desai and B.~Mukhopadhyaya,
  %``Constraints on supersymmetry with light third family from LHC data,''
  JHEP {\bf 1205}, 057 (2012);
  %[arXiv:1111.2830 [hep-ph]].
  %%CITATION = ARXIV:1111.2830;%%
  H.~K.~Dreiner, M.~Kramer and J.~Tattersall,
  %``How low can SUSY go? Matching, monojets and compressed spectra,''
  Europhys.\ Lett.\  {\bf 99}, 61001 (2012);
  %[arXiv:1207.1613 [hep-ph]].
  %%CITATION = ARXIV:1207.1613;%%
  A.~Choudhury and A.~Datta,
  %``New limits on top squark NLSP from LHC 4.7 $fb^{-1}$ data,''
  Mod.\ Phys.\ Lett.\ A {\bf 27}, 1250188 (2012);
  %[arXiv:1207.1846 [hep-ph]].
  %%CITATION = ARXIV:1207.1846;%%
  R.~Mahbubani, M.~Papucci, G.~Perez, J.~T.~Ruderman and A.~Weiler,
  %``Light Nondegenerate Squarks at the LHC,''
  Phys.\ Rev.\ Lett.\  {\bf 110}, no. 15, 151804 (2013);
  %[arXiv:1212.3328 [hep-ph]].
  %%CITATION = ARXIV:1212.3328;%%
  K.~Krizka, A.~Kumar and D.~E.~Morrissey,
  %``Very Light Scalar Top Quarks at the LHC,''
  arXiv:1212.4856 [hep-ph];
  %%CITATION = ARXIV:1212.4856;%%
  C.~Han, K.~-i.~Hikasa, L.~Wu, J.~M.~Yang and Y.~Zhang,
  %``Current experimental bounds on stop mass in natural SUSY,''
  JHEP {\bf 1310}, 216 (2013);
  %[arXiv:1308.5307 [hep-ph]].
  %%CITATION = ARXIV:1308.5307;%%
  J.~S.~Kim, K.~Rolbiecki, K.~Sakurai and J.~Tattersall,
  %```Stop' that ambulance! New physics at the LHC?,''
  arXiv:1406.0858 [hep-ph];  
  %%CITATION = ARXIV:1406.0858;%%
%\cite{Czakon:2014fka}
% \bibitem{Czakon:2014fka} 
  M.~Czakon, A.~Mitov, M.~Papucci, J.~T.~Ruderman and A.~Weiler,
  %``Closing the stop gap,''
  arXiv:1407.1043 [hep-ph].
  %%CITATION = ARXIV:1407.1043;%%


\bibitem{light_stop_lep}
  Z.~Han, A.~Katz, D.~Krohn and M.~Reece,
  %``(Light) Stop Signs,''
  JHEP {\bf 1208}, 083 (2012);
  %[arXiv:1205.5808 [hep-ph]].
  %%CITATION = ARXIV:1205.5808;%%
  E.~L.~Berger, Q.~-H.~Cao, J.~-H.~Yu and H.~Zhang,
  %``Measuring Top Quark Polarization in Top Pair plus Missing Energy Events,''
  Phys.\ Rev.\ Lett.\  {\bf 109}, 152004 (2012);
  %[arXiv:1207.1101 [hep-ph]].
  %%CITATION = ARXIV:1207.1101;%%
  C.~Kilic and B.~Tweedie,
  %``Cornering Light Stops with Dileptonic mT2,''
  JHEP {\bf 1304}, 110 (2013);
  %[arXiv:1211.6106 [hep-ph]].
  %%CITATION = ARXIV:1211.6106;%%
  G.~Belanger, R.~M.~Godbole, L.~Hartgring and I.~Niessen,
  %``Top Polarization in Stop Production at the LHC,''
  JHEP {\bf 1305}, 167 (2013);
  %[arXiv:1212.3526].
  %%CITATION = ARXIV:1212.3526;%%
  E.~L.~Berger, Q.~-H.~Cao, J.~-H.~Yu and H.~Zhang,
  %``Measuring Top-Quark Polarization in Top-Pair + Missing Energy Events,''
  arXiv:1305.7266 [hep-ph];
  %%CITATION = ARXIV:1305.7266;%%
  X.~-Q.~Li, Z.~-G.~Si, K.~Wang, L.~Wang, L.~Zhang and G.~Zhu,
  %``Light Top Squark in Precision Top Quark Sample,''
  Phys.\ Rev.\ D {\bf 89}, 077703 (2014);
  %[arXiv:1311.6874 [hep-ph]].
  %%CITATION = ARXIV:1311.6874;%%
  B.~Dutta, W.~Flanagan, A.~Gurrola, W.~Johns, T.~Kamon, P.~Sheldon, K.~Sinha and K.~Wang {\it et al.},
  %``Probing Compressed Top Squarks at the LHC at 14 TeV,''
  arXiv:1312.1348 [hep-ph].
  %%CITATION = ARXIV:1312.1348;%%

\bibitem{light_stop_had}
  M.~Drees, M.~Hanussek and J.~S.~Kim,
  %``Light Stop Searches at the LHC with Monojet Events,''
  Phys.\ Rev.\ D {\bf 86}, 035024 (2012);
  %[arXiv:1201.5714 [hep-ph]].
  %%CITATION = ARXIV:1201.5714;%%
  D.~S.~M.~Alves, M.~R.~Buckley, P.~J.~Fox, J.~D.~Lykken and C.~-T.~Yu,
  %``Stops and $\not E_T$: The shape of things to come,''
  Phys.\ Rev.\ D {\bf 87}, no. 3, 035016 (2013);
  %[arXiv:1205.5805 [hep-ph]].
  %%CITATION = ARXIV:1205.5805;%%
  B.~Dutta, T.~Kamon, N.~Kolev, K.~Sinha and K.~Wang,
  %``Searching for Top Squarks at the LHC in Fully Hadronic Final State,''
  Phys.\ Rev.\ D {\bf 86}, 075004 (2012);
  % [arXiv:1207.1873 [hep-ph]].
  %%CITATION = ARXIV:1207.1873;%%
  K.~Ghosh, K.~Huitu, J.~Laamanen, L.~Leinonen, K.~Huitu, J.~Laamanen and L.~Leinonen,
  %``Top jets as a probe of degenerate stop-NLSP LSP scenario in the framework of cMSSM,''
  Phys.\ Rev.\ Lett.\  {\bf 110}, 141801 (2013);
  %[arXiv:1207.2429 [hep-ph]].
  %%CITATION = ARXIV:1207.2429;%%
  Z.~-H.~Yu, X.~-J.~Bi, Q.~-S.~Yan and P.~-F.~Yin,
  %``Detecting light stop pairs in coannihilation scenarios at the LHC,''
  Phys.\ Rev.\ D {\bf 87}, 055007 (2013);
  %[arXiv:1211.2997 [hep-ph]].
  %%CITATION = ARXIV:1211.2997;%%
  M.~R.~Buckley, T.~Plehn and M.~J.~Ramsey-Musolf,
  %``Stop on Top,''
  arXiv:1403.2726 [hep-ph].
  %%CITATION = ARXIV:1403.2726;%%

\bibitem{light_stop_prod}
  S.~Kraml and A.~R.~Raklev,
  %``Same-sign top quarks as signature of light stops at the LHC,''
  Phys.\ Rev.\ D {\bf 73}, 075002 (2006);
  % [hep-ph/0512284].
  %%CITATION = HEP-PH/0512284;%%
  S.~Bornhauser, M.~Drees, S.~Grab and J.~S.~Kim,
  %``Light Stop Searches at the LHC in Events with two b-Jets and Missing Energy,''
  Phys.\ Rev.\ D {\bf 83}, 035008 (2011);
  %[arXiv:1011.5508 [hep-ph]].
  %%CITATION = ARXIV:1011.5508;%%
  B.~He, T.~Li and Q.~Shafi,
  %``Impact of LHC Searches on NLSP Top Squark and Gluino Mass,''
  JHEP {\bf 1205}, 148 (2012);
  %[arXiv:1112.4461 [hep-ph]].
  %%CITATION = ARXIV:1112.4461;%%
  C.~-Y.~Chen, A.~Freitas, T.~Han and K.~S.~M.~Lee,
  %``New Physics from the Top at the LHC,''
  JHEP {\bf 1211}, 124 (2012); 
  % [arXiv:1207.4794 [hep-ph]].
  %%CITATION = ARXIV:1207.4794;%%
  C.~Han, K.~-i.~Hikasa, L.~Wu, J.~M.~Yang and Y.~Zhang,
  %``Current experimental bounds on stop mass in natural SUSY,''
  arXiv:1308.5307 [hep-ph].
  %%CITATION = ARXIV:1308.5307;%%


\bibitem{squarkpairLO}  
 G.~L.~Kane and J.~P.~Leveille,
  %``Experimental Constraints On Gluino Masses And Supersymmetric Theories,''
  Phys.\ Lett.\  B {\bf 112}, 227 (1982);  
 P.~R.~Harrison and C.~H.~Llewellyn Smith,
  %``Hadroproduction Of Supersymmetric Particles,''
  Nucl.\ Phys.\  B {\bf 213}, 223 (1983)
  [Erratum-ibid.\  B {\bf 223}, 542 (1983);  
 E.~Reya and D.~P.~Roy,
  %``Supersymmetric Particle Production At P Anti-P Collider Energies,''
  Phys.\ Rev.\  D {\bf 32}, 645 (1985);   
 S.~Dawson, E.~Eichten and C.~Quigg,
  %``Search For Supersymmetric Particles In Hadron - Hadron Collisions,''
  Phys.\ Rev.\  D {\bf 31}, 1581 (1985);
  H.~Baer and X.~Tata,
  %``Component Formulae for Hadroproduction of Left-handed and Right-handed Squarks,''
  Phys.\ Lett.\ B {\bf 160}, 159 (1985).
  %%CITATION = PHLTA,B160,159;%%

\bibitem{squarkpairNLO}
 W.~Beenakker, R.~H\"opker, M.~Spira and P.~M.~Zerwas,
  %``Squark production at the Tevatron,''
  Phys.\ Rev.\ Lett.\  {\bf 74}, 2905 (1995);
  %[arXiv:hep-ph/9412272].
  %%CITATION = PRLTA,74,2905;%%
G.~Bozzi, B.~Fuks and M.~Klasen,
  %``Non-diagonal and mixed squark production at hadron colliders,''
  Phys.\ Rev.\  D {\bf 72}, 035016 (2005).
  %[arXiv:hep-ph/0507073]; 

\bibitem{gluinopairNLO} 
  W.~Beenakker, R.~H\"opker, M.~Spira and P.~M.~Zerwas,
  %``Gluino pair production at the Tevatron,''
  Z.\ Phys.\ C {\bf 69}, 163 (1995).
  %[hep-ph/9505416].
  %%CITATION = HEP-PH/9505416;%%

\bibitem{squarkgluinoNLO}
 W.~Beenakker, R.~H\"opker, M.~Spira and P.~M.~Zerwas,
  %``Squark and gluino production at hadron colliders,''
  Nucl.\ Phys.\  B {\bf 492} (1997) 51.

\bibitem{prospino} 
 available under 
 \url{www.thphys.uni-heidelberg.de/~plehn}

\bibitem{stopNLO}
  W.~Beenakker, M.~Kr\"amer, T.~Plehn, M.~Spira and P.~M.~Zerwas,
  %``Stop production at hadron colliders,''
  Nucl.\ Phys.\  B {\bf 515}, 3 (1998).
  %[arXiv:hep-ph/9710451]; 

\bibitem{squarkEW}
 S.~Bornhauser, M.~Drees, H.~K.~Dreiner and J.~S.~Kim,
  %``Electroweak Contributions to Squark Pair Production at the LHC,''
  Phys.\ Rev.\  D {\bf 76}, 095020 (2007);
  %[arXiv:0709.2544 [hep-ph]];   
 W.~Hollik, M.~Kollar and M.~K.~Trenkel,
  %``Hadronic production of top-squark pairs with electroweak NLO
  %contributions,''
  JHEP {\bf 0802}, 018 (2008);
  %[arXiv:0712.0287 [hep-ph]]; 
 M.~Beccaria, G.~Macorini, L.~Panizzi, F.~M.~Renard and C.~Verzegnassi,
  %``Stop-antistop and sbottom-antisbottom production at LHC: a one-loop search
  %for model parameters dependence,''
  Int.\ J.\ Mod.\ Phys.\  A {\bf 23}, 4779 (2008);
  %[arXiv:0804.1252 [hep-ph]];  
  % \bibitem{Arhrib:2009sb} 
  A.~Arhrib, R.~Benbrik, K.~Cheung and T.~-C.~Yuan,
  %``Higgs boson enhancement effects on squark-pair production at the LHC,''
  JHEP {\bf 1002}, 048 (2010);
%   [arXiv:0911.1820 [hep-ph]].
  %%CITATION = ARXIV:0911.1820;%%
  %7 citations counted in INSPIRE as of 28 Oct 2013
  %[arXiv:0804.1252 [hep-ph]];  
 W.~Hollik, E.~Mirabella and M.~K.~Trenkel,
  %``Electroweak contributions to squark--gluino production at the LHC,''
  JHEP {\bf 0902}, 002 (2009);
  %[arXiv:0810.1044 [hep-ph]]; 
 J.~Germer, W.~Hollik, E.~Mirabella and M.~K.~Trenkel,
  %``Hadronic production of squark-squark pairs: The electroweak
  %contributions,''
  JHEP {\bf 1008}, 023 (2010)
  %[arXiv:1004.2621 [hep-ph]]; 
 J.~Germer, W.~Hollik and E.~Mirabella,
  %``Hadronic production of bottom-squark pairs with electroweak
  %contributions,''
  JHEP {\bf 1105}, 068 (2011); 
  %[arXiv:1103.1258 [hep-ph]].
  W.~Hollik, J.~M.~Lindert and D.~Pagani,
  %``NLO corrections to squark-squark production and decay at the LHC,''
  JHEP {\bf 1303}, 139 (2013);
  W.~Hollik, J.~M.~Lindert and D.~Pagani,
  %``On cascade decays of squarks at the LHC in NLO QCD,''
  Eur.\ Phys.\ J.\ C {\bf 73}, 2410 (2013);
  %[arXiv:1303.0186 [hep-ph]].
  %%CITATION = ARXIV:1303.0186;%%
    J.~Germer, W.~Hollik, J.~M.~Lindert and E.~Mirabella,
  %``Top-squark pair production at the LHC: a complete analysis at next-to-leading order,''
  arXiv:1404.5572 [hep-ph].
  %%CITATION = ARXIV:1404.5572;%%
  
\bibitem{squarkgluinoNNLO}
 U.~Langenfeld and S.~-O.~Moch,
  %``Higher-order soft corrections to squark hadro-production,''
  Phys.\ Lett.\ B {\bf 675}, 210 (2009);
  %[arXiv:0901.0802 [hep-ph]].
  %%CITATION = ARXIV:0901.0802;%%
  U.~Langenfeld, S.~-O.~Moch and T.~Pfoh,
  %``QCD threshold corrections for gluino pair production at hadron colliders,''
  JHEP {\bf 1211}, 070 (2012).
  %[arXiv:1208.4281 [hep-ph]].
  %%CITATION = ARXIV:1208.4281;%%

\bibitem{squarkgluinoResummed}
 W.~Beenakker, S.~Brensing, M.~Kr\"amer, A.~Kulesza, E.~Laenen and I.~Niessen,
  %``Soft-gluon resummation for squark and gluino hadroproduction,''
  JHEP {\bf 0912}, 041 (2009);
  %[arXiv:0909.4418 [hep-ph]].
 J.~Debove, B.~Fuks and M.~Klasen,
  %``Joint Resummation for Gaugino Pair Production at Hadron Colliders,''
  Nucl.\ Phys.\ B {\bf 849}, 64 (2011);
  %[arXiv:1102.4422 [hep-ph]].
  %%CITATION = ARXIV:1102.4422;%%
 A.~Kulesza and L.~Motyka,
  %``Threshold resummation for squark-antisquark and gluino-pair production at
  %the LHC,''
  Phys.\ Rev.\ Lett.\  {\bf 102}, 111802 (2009);
  %[arXiv:0807.2405 [hep-ph]];
 A.~Kulesza and L.~Motyka,
  %``Soft gluon resummation for the production of gluino-gluino and
  %squark-antisquark pairs at the LHC,''
  Phys.\ Rev.\  D {\bf 80}, 095004 (2009);
  %[arXiv:0905.4749 [hep-ph]];   
 M.~Beneke, P.~Falgari and C.~Schwinn,
  %``Threshold resummation for pair production of coloured heavy (s)particles at
  %hadron colliders,''
  Nucl.\ Phys.\  B {\bf 842}, 414 (2011);  P.~Falgari, C.~Schwinn and C.~Wever,
  %``NLL soft and Coulomb resummation for squark and gluino production at the LHC,''
  JHEP {\bf 1206}, 052 (2012); 
  %[arXiv:1007.5414 [hep-ph]];
 M.~R.~Kauth, J.~H.~K\"uhn, P.~Marquard and M.~Steinhauser,
  %``Gluino Pair Production at the LHC: The Threshold,''
  Nucl.\ Phys.\ B {\bf 857}, 28 (2012).
  %[arXiv:1108.0361 [hep-ph]].
  %%CITATION = ARXIV:1108.0361;%%

\bibitem{Gavin:2013kga} 
  R.~Gavin, C.~Hangst, M.~Kr\"amer, M.~M\"uhlleitner, M.~Pellen, E.~Popenda and M.~Spira,
  %``Matching Squark Pair Production at NLO with Parton Showers,''
  arXiv:1305.4061 [hep-ph].
  %%CITATION = ARXIV:1305.4061;%%

\bibitem{Binoth:2011xi} 
  T.~Binoth, D.~Gon\c{c}alves-Netto, D.~L\'opez-Val, K.~Mawatari, T.~Plehn and I.~Wigmore,
  %``Automized Squark-Neutralino Production to Next-to-Leading Order,''
  Phys.\ Rev.\ D {\bf 84}, 075005 (2011).
  %[arXiv:1108.1250 [hep-ph]].
  %%CITATION = ARXIV:1108.1250;%%
  %19 citations counted in INSPIRE as of 02 Sep 2013

\bibitem{GoncalvesNetto:2012yt} 
  D.~Gon\c{c}alves-Netto, D.~L\'opez-Val, K.~Mawatari, T.~Plehn and I.~Wigmore,
  %``Automated Squark and Gluino Production to Next-to-Leading Order,''
  Phys.\ Rev.\ D {\bf 87}, 014002 (2013).
  %[arXiv:1211.0286 [hep-ph]].
  %%CITATION = ARXIV:1211.0286;%%

\bibitem{madgolem_nonsusy} 
  D.~Gon\c{c}alves-Netto, D.~L\'opez-Val, K.~Mawatari, T.~Plehn and I.~Wigmore,
  %``Sgluon Pair Production to Next-to-Leading Order,''
  Phys.\ Rev.\ D {\bf 85}, 114024 (2012);
  %[arXiv:1203.6358 [hep-ph]].
  %%CITATION = ARXIV:1203.6358;%%
  D.~Gon\c{c}alves-Netto, D.~L\'opez-Val, K.~Mawatari, I.~Wigmore and T.~Plehn,
  %``Looking for leptogluons,''
  Phys.\ Rev.\ D {\bf 87}, 094023 (2013).
  %[arXiv:1303.0845 [hep-ph]].
  %%CITATION = ARXIV:1303.0845;%%

\bibitem{mg4} 
 J.~Alwall {\it et al.},
  %``MadGraph/MadEvent v4: The New Web Generation,''
  JHEP {\bf 0709}, 028 (2007).
  %[arXiv:0706.2334 [hep-ph]].
  %%CITATION = JHEPA,0709,028;%%

\bibitem{mg5}
  J.~Alwall, M.~Herquet, F.~Maltoni, O.~Mattelaer and T.~Stelzer,
  %``MadGraph 5 : Going Beyond,''
  JHEP {\bf 1106}, 128 (2011).
  %[arXiv:1106.0522 [hep-ph]].
  %%CITATION = ARXIV:1106.0522;%%

%\bibitem{helas}
%  H.~Murayama, I.~Watanabe, K.~Hagiwara,
%  %``HELAS: HELicity amplitude subroutines for Feynman diagram evaluations,''
%  prepring KEK-91-11.
  
\bibitem{qgraf} 
 P.~Nogueira,
  J. Comp. Phys. {\bf 105}, 279 (1993).

\bibitem{golem}
 T.~Binoth, J.~P~.Guillet, G.~Heinrich, E.~Pilon and C.~Schubert,
  %``An Algebraic/numerical formalism for one-loop multi-leg amplitudes,''
  JHEP {\bf 0510} (2005) 015;
  %[hep-ph/0504267].
  %%CITATION = HEP-PH/0504267;%%
 G.~Cullen, N.~Greiner, A.~Guffanti, J.~P.~Guillet, G.~Heinrich, S.~Karg, N.~Kauer, T.~Kleinschmidt {\it et al.},
  %``Recent Progress in the Golem Project,''
  Nucl.\ Phys.\ Proc.\ Suppl.\  {\bf 205-206 } (2010)  67-73.
  %[arXiv:1007.3580 [hep-ph]].

\bibitem{golem_lib}
 T.~Binoth, J.~P.~Guillet, G.~Heinrich, E.~Pilon and T.~Reiter,
  %``Golem95: A Numerical program to calculate one-loop tensor integrals with up
  %to six external legs,''
  Comput.\ Phys.\ Commun.\  {\bf 180}, 2317 (2009);
  %[arXiv:0810.0992 [hep-ph]]; 
 G.~Cullen, J.~P.~Guillet, G.~Heinrich, T.~Kleinschmidt, E.~Pilon, T.~Reiter and M.~Rodgers,
 Comput.\ Phys.\ Commun.\  {\bf 182}, 2276 (2011).
  %``Golem95C: A library for one-loop integrals with complex masses,''
%   arXiv:1101.5595 [hep-ph].
  %%CITATION = ARXIV:1101.5595;%%

\bibitem{catani_seymour}
 S.~Catani and M.~H.~Seymour,
  %``A General algorithm for calculating jet cross-sections in NLO QCD,''
  Nucl.\ Phys.\  B {\bf 485}, 291 (1997)
  [Erratum-ibid.\  B {\bf 510}, 503 (1998)];
  %[arXiv:hep-ph/9605323]; 
 S.~Catani, S.~Dittmaier, M.~H.~Seymour and Z.~Trocsanyi,
  Nucl.\ Phys.\  B {\bf 627}, 189 (2002).
  %[arXiv:hep-ph/0201036].

\bibitem{on-shell}
 for some more details see \eg 
 T.~Plehn, C.~Weydert,
  %``Charged Higgs production with a top in MC@NLO,''
  PoS {\bf CHARGED2010}, 026 (2010).
  %[arXiv:1012.3761 [hep-ph]].

\bibitem{maddipole} 
 R.~Frederix, T.~Gehrmann and N.~Greiner,
  %``Automation of the Dipole Subtraction Method in MadGraph/MadEvent,''
  JHEP {\bf 0809}, 122 (2008);
  %[arXiv:0808.2128 [hep-ph]]; 
  %``Integrated dipoles with MadDipole in the MadGraph framework,''
  and
  JHEP {\bf 1006}, 086 (2010).
  %[arXiv:1004.2905 [hep-ph]].

\bibitem{alpha} 
 S.~Frixione, Z.~Kunszt and A.~Signer,
  %``Three jet cross-sections to next-to-leading order,''
  Nucl.\ Phys.\  B {\bf 467}, 399 (1996);
  %[arXiv:hep-ph/9512328]; 
  Z.~Nagy and Z.~Trocsanyi,
  %``Next-to-leading order calculation of four jet observables in electron
  %positron annihilation,''
  Phys.\ Rev.\  D {\bf 59}, 014020 (1999)
  [Erratum-ibid.\  D {\bf 62}, 099902 (2000)].
  %[arXiv:hep-ph/9806317].

\bibitem{Martin:1993yx}
 S.~P.~Martin and M.~T.~Vaughn,
  %``Regularization dependence of running couplings in softly broken
  %supersymmetry,''
  Phys.\ Lett.\  B {\bf 318}, 331 (1993);
  %[arXiv:hep-ph/9308222].
  %%CITATION = PHLTA,B318,331;%%
  A.~Signer, D.~St\"ockinger,
  %``Factorization and regularization by dimensional reduction,''
  Phys.\ Lett.\  {\bf B626}, 127-138 (2005);
  %[hep-ph/0508203].
  %A.~Signer, D.~Stockinger,
  and
  %``Using Dimensional Reduction for Hadronic Collisions,''
  Nucl.\ Phys.\  {\bf B808}, 88-120 (2009).
  %[arXiv:0807.4424 [hep-ph]].

\bibitem{cteq}
 J.~Pumplin, D.~R.~Stump, J.~Huston, H.~L.~Lai, P.~Nadolsky and W.~K.~Tung,
  %``New generation of parton distributions with uncertainties from global  QCD
  %analysis,''
  JHEP {\bf 0207}, 012 (2002).
  %[arXiv:hep-ph/0201195];

\bibitem{Ball:2011mu} 
  R.~D.~Ball, V.~Bertone, F.~Cerutti, L.~Del Debbio, S.~Forte, A.~Guffanti, J.~I.~Latorre and J.~Rojo {\it et al.},
  %``Impact of Heavy Quark Masses on Parton Distributions and LHC Phenomenology,''
  Nucl.\ Phys.\ B {\bf 849}, 296 (2011).
  %[arXiv:1101.1300 [hep-ph]].
  %%CITATION = ARXIV:1101.1300;%%

\bibitem{Berger:2008cq} 
  C.~F.~Berger, J.~S.~Gainer, J.~L.~Hewett and T.~G.~Rizzo,
  %``Supersymmetry Without Prejudice,''
  JHEP {\bf 0902}, 023 (2009).
  %[arXiv:0812.0980 [hep-ph]].
  %%CITATION = ARXIV:0812.0980;%%

\bibitem{Cahill-Rowley:2013gca} 
  M.~W.~Cahill-Rowley, J.~L.~Hewett, A.~Ismail, M.~E.~Peskin and T.~G.~Rizzo,
  %``pMSSM Benchmark Models for Snowmass 2013,''
  arXiv:1305.2419 [hep-ph].
  %%CITATION = ARXIV:1305.2419;%%

\bibitem{sps} 
  S.~S.~AbdusSalam, B.~C.~Allanach, H.~K.~Dreiner, J.~Ellis, U.~Ellwanger, J.~Gunion, S.~Heinemeyer and M.~Kr\"amer {\it et al.},
  %``Benchmark Models, Planes, Lines and Points for Future SUSY Searches at the LHC,''
  Eur.\ Phys.\ J.\ C {\bf 71}, 1835 (2011).
  %[arXiv:1109.3859 [hep-ph]].
  %%CITATION = ARXIV:1109.3859;%%

\bibitem{Allanach:2001kg} 
  B.~C.~Allanach,
  %``SOFTSUSY: a program for calculating supersymmetric spectra,''
  Comput.\ Phys.\ Commun.\  {\bf 143}, 305 (2002).
  %[hep-ph/0104145].
  %%CITATION = HEP-PH/0104145;%%

\bibitem{pythia}
 T.~Sj\"ostrand, S.~Mrenna and P.~Z.~Skands,
  %``PYTHIA 6.4 Physics and Manual,''
  JHEP {\bf 0605}, 026 (2006).
  %[hep-ph/0603175].
  %%CITATION = HEP-PH/0603175;%%

\bibitem{mlm}
 M.~L.~Mangano, M.~Moretti, and R.~Pittau,
  %``Multijet matrix elements and shower evolution in hadronic collisions:  W b
  %anti-b + (n)jets as a case study,''
  Nucl.\ Phys.\  B {\bf 632}, 343 (2002).
  %[arXiv:hep-ph/0108069].
  %%CITATION = NUPHA,B632,343;%%

\bibitem{madgraph_merging}
 J.~Alwall, S.~de Visscher, F.~Maltoni,
  %``QCD radiation in the production of heavy colored particles at the LHC,''
  JHEP {\bf 0902 },  017 (2009).
  %[arXiv:0810.5350 [hep-ph]].
  %%CITATION = JHEPA,0902,017;%%  

\bibitem{qcd_radiation}
 T.~Plehn, D.~Rainwater, P.~Z.~Skands,
  %``Squark and gluino production with jets,''
  Phys.\ Lett.\  {\bf B645}, 217-221 (2007);
  %[hep-ph/0510144].
  %%CITATION = PHLTA,B645,217;%%
 T.~Plehn and T.~M.~P.~Tait,
  %``Seeking Sgluons,''
  J.\ Phys.\ G {\bf 36}, 075001 (2009).
  %[arXiv:0810.3919 [hep-ph]].
  %%CITATION = JPHGB,G36,075001;%%

\bibitem{autofocus}
 C.~Englert, T.~Plehn, P.~Schichtel and S.~Schumann,
  %``Jets plus Missing Energy with an Autofocus,''
  Phys.\ Rev.\ D {\bf 83}, 095009 (2011).
  %[arXiv:1102.4615 [hep-ph]].
  %%CITATION = ARXIV:1102.4615;%%

  
\bibitem{4vs5higgs} 
  T.~Plehn,
  %``Charged Higgs boson production in bottom gluon fusion,''
  Phys.\ Rev.\ D {\bf 67}, 014018 (2003);
  %[hep-ph/0206121].
  %%CITATION = HEP-PH/0206121;%%
  %172 citations counted in INSPIRE as of 04 Apr 2014 
  E.~Boos and T.~Plehn,
  %``Higgs boson production induced by bottom quarks,''
  Phys.\ Rev.\ D {\bf 69}, 094005 (2004). 
  %[hep-ph/0304034].
  %%CITATION = HEP-PH/0304034;%%
  %78 citations counted in INSPIRE as of 04 Apr 2014  
  
  \bibitem{acot} 
  J.~C.~Collins and W.~-K.~Tung,
  %``Calculating Heavy Quark Distributions,''
  Nucl.\ Phys.\ B {\bf 278}, 934 (1986); 
  %%CITATION = NUPHA,B278,934;%%
  R.~M.~Barnett, H.~E.~Haber and D.~E.~Soper,
  %``Ultraheavy Particle Production from Heavy Partons at Hadron Colliders,''
  Nucl.\ Phys.\ B {\bf 306}, 697 (1988).
  %%CITATION = NUPHA,B306,697;%%
  %160 citations counted in INSPIRE as of 08 Apr 2014  
  F.~I.~Olness and W.~-K.~Tung,
  %``When Is a Heavy Quark Not a Parton? Charged Higgs Production and Heavy Quark Mass Effects in the QCD Based Parton Model,''
  Nucl.\ Phys.\ B {\bf 308}, 813 (1988).
  %%CITATION = NUPHA,B308,813;%%
  M.~A.~G.~Aivazis, J.~C.~Collins, F.~I.~Olness and W.~-K.~Tung,
  %``Leptoproduction of heavy quarks. 2. A Unified QCD formulation of charged and neutral current processes from fixed target to collider energies,''
  Phys.\ Rev.\ D {\bf 50}, 3102 (1994); 
  %[hep-ph/9312319].
  %%CITATION = HEP-PH/9312319;%%
  J.~C.~Collins,
  %``Hard scattering factorization with heavy quarks: A General treatment,''
  Phys.\ Rev.\ D {\bf 58}, 094002 (1998);
  %[hep-ph/9806259].
  %%CITATION = HEP-PH/9806259;%%
  M.~Kr\"amer, 1, F.~I.~Olness and D.~E.~Soper,
  %``Treatment of heavy quarks in deeply inelastic scattering,''
  Phys.\ Rev.\ D {\bf 62}, 096007 (2000).
  %[hep-ph/0003035].
  %%CITATION = HEP-PH/0003035;%%
 

\bibitem{Maltoni:2012pa} 
  for a recent overview see \eg
  F.~Maltoni, G.~Ridolfi and M.~Ubiali,
  %``b-initiated processes at the LHC: a reappraisal,''
  JHEP {\bf 1207}, 022 (2012)
  [Erratum-ibid.\  {\bf 1304}, 095 (2013)].
  %[arXiv:1203.6393 [hep-ph]].
  %%CITATION = ARXIV:1203.6393;%%

\bibitem{bbhiggs}
  F.~Maltoni, Z.~Sullivan and S.~Willenbrock,
  %``Higgs-boson production via bottom-quark fusion,''
  Phys.\ Rev.\ D {\bf 67}, 093005 (2003); 
  %[hep-ph/0301033].
  %%CITATION = HEP-PH/0301033;%%
  S.~Dawson, C.~B.~Jackson, L.~Reina and D.~Wackeroth,
  %``Higgs production in association with bottom quarks at hadron colliders,''
  Mod.\ Phys.\ Lett.\ A {\bf 21}, 89 (2006);
  %[hep-ph/0508293].
  %%CITATION = HEP-PH/0508293;%%
  R.~V.~Harlander and W.~B.~Kilgore,
  %``Next-to-next-to-leading order Higgs production at hadron colliders,''
  Phys.\ Rev.\ Lett.\  {\bf 88}, 201801 (2002).
  %[hep-ph/0201206].

\bibitem{early-squark}
  W.~Beenakker, R.~H\"opker and P.~M.~Zerwas,
  %``SUSY QCD decays of squarks and gluinos,''
  Phys.\ Lett.\ B {\bf 378}, 159 (1996);
  %[hep-ph/9602378];
  %%CITATION = HEP-PH/9602378;%%
  S.~Kraml, H.~Eberl, A.~Bartl, W.~Majerotto and W.~Porod,
  %``SUSY QCD corrections to scalar quark decays into charginos and neutralinos,''
  Phys.\ Lett.\ B {\bf 386}, 175 (1996);
  %[hep-ph/9605412];
  %%CITATION = HEP-PH/9605412;%%
  A.~Djouadi, W.~Hollik and C.~J\"unger,
  %``QCD corrections to scalar quark decays,''
  Phys.\ Rev.\ D {\bf 55}, 6975 (1997);
  %[hep-ph/9609419].
  %%CITATION = HEP-PH/9609419;%%
  W.~Beenakker, R.~H\"opker, T.~Plehn and P.~M.~Zerwas,
  %``Stop decays in SUSY QCD,''
  Z.\ Phys.\ C {\bf 75}, 349 (1997);
  %[hep-ph/9610313];
  %%CITATION = HEP-PH/9610313;%%
  J.~Guasch, J.~Sol\`a and W.~Hollik,
  %``Yukawa coupling corrections to scalar quark decays,''
  Phys.\ Lett.\ B {\bf 437}, 88 (1998);
  %[hep-ph/9802329];
  A.~Bartl, H.~Eberl, K.~Hidaka, S.~Kraml, W.~Majerotto, W.~Porod and Y.~Yamada,
  %``SUSY - QCD corrections to top and bottom squark decays into Higgs bosons,''
  Phys.\ Rev.\ D {\bf 59}, 115007 (1999);
  %[hep-ph/9806299];
  %%CITATION = HEP-PH/9806299;%%
  M.~M\"uhlleitner, A.~Djouadi and Y.~Mambrini,
  %``SDECAY: A Fortran code for the decays of the supersymmetric particles in the MSSM,''
  Comput.\ Phys.\ Commun.\  {\bf 168}, 46 (2005);
  %[hep-ph/0311167];
  %%CITATION = HEP-PH/0311167;%%
  L.~G.~Jin and C.~S.~Li,
  %``Supersymmetric electroweak corrections to sbottom decay into lighter stop and charged Higgs boson,''
  Phys.\ Rev.\ D {\bf 65}, 035007 (2002);
  %[hep-ph/0106253];
  %%CITATION = HEP-PH/0106253;%%
  J.~Guasch, W.~Hollik and J.~Sol\`a,
  %``Full electroweak one loop radiative corrections to squark decays in the MSSM,''
  Phys.\ Lett.\ B {\bf 510}, 211 (2001);
  %[hep-ph/0101086];
  %%CITATION = HEP-PH/0101086;%%
  J.~Guasch, W.~Hollik and J.~Sol\`a,
  %``Fermionic decays of sfermions: A Complete discussion at one loop order,''
  JHEP {\bf 0210}, 040 (2002);
  %[hep-ph/0207364];
  %%CITATION = HEP-PH/0207364;%%
  A.~Arhrib and R.~Benbrik,
  %``Third generation sfermions decays into $Z$ and $W$ gauge bosons: Full one-loop analysis,''
  Phys.\ Rev.\ D {\bf 71}, 095001 (2005).
  %[hep-ph/0412349].
  %%CITATION = HEP-PH/0412349;%%

\bibitem{sfermionsector}
  W.~Hollik and H.~Rzehak,
  %``The Sfermion mass spectrum of the MSSM at the one loop level,''
  Eur.\ Phys.\ J.\ C {\bf 32}, 127 (2003).
  %[hep-ph/0305328];
 
\bibitem{thesis}
 M.~A.~Diaz,
  %``Diagonalization of coupled scalars and its application to the supersymmetric neutral Higgs sector,''
  hep-ph/9705471;
  %%CITATION = HEP-PH/9705471;%%
 T.~Plehn and W.~Beenakker,
  %``Stop mixing in the MSSM,''
  In *Barcelona 1997, Quantum effects in the minimal supersymmetric standard model* 244-249.
  
\bibitem{renorm-higgs}
  Q.~Li, L.~G.~Jin and C.~S.~Li,
  %``Supersymmetric electroweak corrections to heavier top squark decay into lighter top squark and neutral Higgs boson,''
  Phys.\ Rev.\ D {\bf 66}, 115008 (2002);
  %[hep-ph/0207363];
  %%CITATION = HEP-PH/0207363;%%
  %%CITATION = HEP-PH/0305328;%%
  C.~Weber, K.~Kovarik, H.~Eberl and W.~Majerotto,
  %``Complete one-loop corrections to decays of charged and CP-even neutral Higgs bosons into sfermions,''
  Nucl.\ Phys.\ B {\bf 776}, 138 (2007).
  %[hep-ph/0701134];
  %%CITATION = HEP-PH/0701134;%%

\bibitem{ren-real}
  A.~Djouadi, P.~Gambino, S.~Heinemeyer, W.~Hollik, C.~J\"unger and G.~Weiglein,
  %``Leading QCD corrections to scalar quark contributions to electroweak precision observables,''
  Phys.\ Rev.\ D {\bf 57}, 4179 (1998);
  %[hep-ph/9710438].
  %%CITATION = HEP-PH/9710438;%%
  S.~Heinemeyer, W.~Hollik, H.~Rzehak and G.~Weiglein,
  %``High-precision predictions for the MSSM Higgs sector at O(alpha(b) alpha(s)),''
  Eur.\ Phys.\ J.\ C {\bf 39}, 465 (2005);
  %[hep-ph/0411114].
  %%CITATION = HEP-PH/0411114;%%
  N.~Baro and F.~Boudjema,
  %``Automatised full one-loop renormalisation of the MSSM II: The chargino-neutralino sector, the sfermion sector and some applications,''
  Phys.\ Rev.\ D {\bf 80}, 076010 (2009).
  %[arXiv:0906.1665 [hep-ph]].
  %%CITATION = ARXIV:0906.1665;%%

\bibitem{Heinemeyer:2010mm} 
  S.~Heinemeyer, H.~Rzehak and C.~Schappacher,
  %``Proposals for Bottom Quark/Squark Renormalization in the Complex MSSM,''
  Phys.\ Rev.\ D {\bf 82}, 075010 (2010).
  %[arXiv:1007.0689 [hep-ph]].
  %%CITATION = ARXIV:1007.0689;%%

\bibitem{ren-complex}
  S.~Heinemeyer, W.~Hollik, H.~Rzehak and G.~Weiglein,
  %``The Higgs sector of the complex MSSM at two-loop order: QCD contributions,''
  Phys.\ Lett.\ B {\bf 652}, 300 (2007);
  %[arXiv:0705.0746 [hep-ph]].
  %%CITATION = ARXIV:0705.0746;%%
  T.~Fritzsche, S.~Heinemeyer, H.~Rzehak and C.~Schappacher,
  %``Heavy Scalar Top Quark Decays in the Complex MSSM: A Full One-Loop Analysis,''
  Phys.\ Rev.\ D {\bf 86}, 035014 (2012).
  %[arXiv:1111.7289 [hep-ph]].
  %%CITATION = ARXIV:1111.7289;%%

\bibitem{Hollik:2002mv} 
  W.~Hollik, E.~Kraus, M.~Roth, C.~Rupp, K.~Sibold and D.~St\"ockinger,
  %``Renormalization of the minimal supersymmetric standard model,''
  Nucl.\ Phys.\ B {\bf 639}, 3 (2002).
  %[hep-ph/0204350].
  %%CITATION = HEP-PH/0204350;%%

\bibitem{Berge:2007dz} 
  S.~Berge, W.~Hollik, W.~M.~Mosle and D.~Wackeroth,
  %``SUSY QCD one-loop effects in (un)polarized top-pair production at hadron colliders,''
  Phys.\ Rev.\ D {\bf 76}, 034016 (2007).
  %[hep-ph/0703016 [HEP-PH]].
  %%CITATION = HEP-PH/0703016;%%

\bibitem{decoup}
  J.~C.~Collins, F.~Wilczek and A.~Zee,
  %``Low-Energy Manifestations Of Heavy Particles: Application To The Neutral
  %Current,''
  Phys.\ Rev.\  D {\bf 18}, 242 (1978).
  %%CITATION = PHRVA,D18,242;%%

\bibitem{oneloop}
  A.~van Hameren,
  %``OneLOop: for the evaluation of one-loop scalar functions,''
  Comput.\ Phys.\ Commun.\  {\bf 182}, 2427 (2011).
  %[arXiv:1007.4716 [hep-ph]].

\bibitem{Fritzsche:2002bi} 
  T.~Fritzsche and W.~Hollik,
  %``Complete one loop corrections to the mass spectrum of charginos and neutralinos in the MSSM,''
  Eur.\ Phys.\ J.\ C {\bf 24}, 619 (2002).
  %[hep-ph/0203159].
  %%CITATION = HEP-PH/0203159;%%

\end{thebibliography}
\end{document}